# Investigating the dynamics of point helical vortices on a rotating sphere to model tropical cyclones

Sergey G. Chefranov [1);a)], Igor I. Mokhov [1);2);3);b)] and Alexander G. Chefranov[4); c)]

1) A.M. Obukhov Institute of Atmospheric Physics RAS, Moscow 119017, Russia
2) M. V. Lomonosov Moscow State University, Moscow 119991, Russia
3) Moscow Institute for Physics and Technology, Dolgoprudny 141700, Russia
4) Eastern Mediterranean University, Famagusta 99628, North Cyprus

a) Author to whom correspondence should be addressed: schefranov@mail.ru

b) mokhov@ifaran.ru ; c) Alexander.chefranov@emu.edu.tr

## Abstract

A general exact weak solution to the nonlinear equation of the conservation of the absolute vorticity in a thin layer of an incompressible medium on a rotating sphere is proposed. It takes into account the helicity of the point vortices and the non-uniformity of the depth of the layer. This is used to develop a model of the observed interactions of spiral atmospheric vortices. The fusion of two point helical vortices (HVs) on the rotating sphere is considered. We also analyze the prognostic applicability of the dynamics of the HVs for modeling the abrupt changes observed in the trajectories of tropical cyclones and their landfall in comparison with the traditional approach. The analytical condition for chiral symmetry violation related to the direction of the movement of the center of a helical cyclone is obtained.

## I. Introduction

Vortical motions have been observed in many natural and artificial systems. The corresponding localized formation of vortices provides an example of the self-organizing non-equilibrium system considered in Ref. 1 in connection with the problem of evolution of typhoons, long before the idea of synergetic dissipative structures was proposed.[2] Indeed, the thesis stated in Ref. 1—"If a vortex by its own action can increase its intensity to a higher degree than has ever existed in the surrounding medium by taking energy from it, we can regard it as having animate growth"—provides an example of a phenomenon of selection, like mechanical or electro-magnetic resonance. Thus, the first and second laws of thermodynamics are not violated in the observed growth in the intensity of the vortex, like in other open and irreversible dissipative systems that are considered in synergetic theory.[2]

Research on the dynamics of localized vortices and their interactions, which lead to the concentration of vorticity observed in nature and in laboratory experiments, has fundamental theoretical and practical importance.[1, 3-23]

It is also well known that the introduction of such idealized hydrodynamical objects as a point vortex, and a point source or sink for investigating fluid flow are useful for solving a number of problems in hydrodynamics, geophysics, the physics of magnetized plasma, and theories of superfluids and superconductivity.[3-7, 16-23]

Many studies have been devoted to point vortices, and some of the earliest have been reviewed in Refs. 3–5. Research has also investigated the motion of free point sinks and sources.[6, 7, 18]

A characteristic feature of observed vortices is the convergent radial motion of the medium at their base, such as in the case of sink vortices in bathtubs, whirlwinds, tornadoes, polar meso-cyclones and typhoons.[1, 8-15, 18] Convergence in these vortices can be simulated by means of sinks, the physical nature of which is typically not considered. For example, in case of typhoons, the convection above the surface of the ocean is modeled by a sink–source system when the sink is located on the lower level and the source on the upper level of the troposphere.[6] The problem is then reduced to one of two-dimensional (2D) hydrodynamics by using a simple model when a combination of a point vortex and a sink is considered.[4, 6, 7, 18] It is also called a point helical vortex (HV) due to the form of its stream function.[4, 18]

The physical mechanism of the stability of the observed helical motion when modeled by an HV defines a constrained set of variants such that the point vortex with a given circulation on a plane or a spherical surface is consistent only with a point sink, or vice versa only with the source. On the contrary, the formal mathematical formulation allows for any such variant.[4, 7] For example, the authors of Ref. 7 considered the Hamiltonian dynamics of an HV on a plane. They claimed that such symmetry in the choice of the combination of a vortex with a sink/source is not violated for either of two possible directions of rotation of the vortex. However, this symmetry is violated when accounting for the rotation of the system as a whole.[24, 25] References 24 and 25 obtained violations of cyclone–anti-cyclone symmetry when accounting for the rotation of the system as a whole and considering the near-surface Ekman friction. It was shown that cyclonic vortical rotation (complying with the direction of rotation of the Earth) can be related only with a sink, but only if the speed of rotation of a fluid is greater than that of the Earth. If such an excess of velocity is absent, only a combination of anti-cyclonic rotation with a radial motion from the axis of rotation is stable. This may be modeled as point source similar to a convective cell.[6] Adding circulation contributes to the stability of pure radial motion in other cases as well.[26-28] The resulting spiral motion of the medium can be modeled by an HV due to its stability.

In this study, we consider the dynamics of an HV on a rotating sphere. The exact solutions obtained are evaluated from the point of view of the possibility of describing features of the dynamics of tropical cyclones (TCs).[8-10, 29-39]

In particular, the effect of the mutual convergence of two TCs—the Fujiwhara effect—has been observed,[1, 9] and has theoretical as well as practical importance for hydrodynamics and turbulence theory. Explaining the Fujiwhara effect in the framework of point non-helical vortex theory in an ideal liquid is impossible because the distance between the monopoles of point vortices is known to be invariant.[3-5] Understanding the mechanism of two vortices approaching each other in different physical systems continues to be the subject of active research in the area.[40-43]

We also consider the effects of attraction of an HV to a boundary in this study. The method of mirror reflections was used in Ref. 18 to describe this phenomenon. Moreover, we assess the applicability of the proposed approach for modeling the interaction between the TC and the coastline in comparison with traditional methods of predicting the trajectories of the TC.[29-39]

Traditional methods can be used to identify general patterns of the trajectories of TCs, against the background of their complex dynamics that are associated with the convective processes of their formation, by considering the regional features of atmospheric circulation over the ocean in tropical latitudes. The TC shifts to the west in the tropical zone of trade winds, with a deviation to higher latitudes due to the influence of the Coriolis force. Upon reaching the latitudinal boundary of the Hadley meridional cell and entering the zone of westerly winds within the Ferrell meridional cell, the trajectory of the TC turns sharply to the east. Moreover, the remote influence of land on the trajectory of the TC has not yet been considered, to the best of the authors' knowledge. Although known process models can be built to predict the recurvature and landfall of the TC, they generally fall short of capturing the intricacies of the underlying mechanisms of such movements.[29-39]

In this paper, we evaluate in particular the possibility of interactions between the TC and the mainland within the framework of the proposed approach. The TC can be attracted to the boundaries of land and the ocean, like the well-known attraction of smaller, helical atmospheric vortices to a solid boundary (see Ref. 18).

The strong interactions involving large-scale atmospheric vortices and waves are traditionally modeled based on the known absolute vorticity conservation equation (AVCE) of a barotropic incompressible fluid in a thin layer (with a non-constant depth in the general case) on a rotating sphere.[5, 19, 44] To solve the problem according to Ref. 44, it is necessary to consider the system of ordinary differential equations (ODEs), which is not closed and has infinite dimensions, to obtain the amplitudes of modes of the interacting waves.

In this study, we generalize an earlier theory[19] to obtain a finite-dimensional and closed system of ODEs that can fully account for the non-linearity of the AVCE as well as the strong interactions of localized singular helical vortices. We also consider the non-uniformity of the depth of the thin layer of fluid on the rotating sphere and the helicity of point vortices to complement the theory.[19] In contrast to past theoretical discussions of point non-helical vortices on a rotating sphere, we obtain an exact weak solution to the AVCE while using singular representation for the absolute value of vorticity, and not the local vorticity, as is usually the case.

The remainder of this paper is structured as follows: We consider the dynamics of point helical vortices on a rotating sphere in Section II by using the proposed exact solution of the AVCE for a rotating sphere in a form that is more common than the one proposed in Ref. 19. We also show the advantages of using singular vortices in comparison with the theory of interactions between non-linear modes of waves.[44] Section III details the interaction of a single HV with a boundary and Section IV discusses the interactions between two HVs. Appendix A presents the general form of the dynamics of an arbitrary system of an HV on a rotating sphere, and Appendix B derives the dynamical system of an HV on the beta plane.

## II. Dynamics equations of helical point vortices

Consider the vortical dynamics of a thin layer of an ideal incompressible liquid in the spherical coordinates $(r;\theta;\varphi)$ on the surface of a rotating sphere of radius R.[5, 19] It is assumed that $r - R \le H << R$, and that variations in the radial coordinate can be neglected when it is constant, i.e., $r \approx R$.[5] The AVCE of a thin layer of an incompressible fluid with a variable depth $H = H(\theta;\varphi;t)$ is as follows:[5]

$$\frac{\partial(\omega/H)}{\partial t} + \frac{V_\theta}{R}\frac{\partial(\omega/H)}{\partial\theta} + \frac{V_\varphi}{R\sin\theta}\frac{\partial(\omega/H)}{\partial\varphi} = 0 \tag{1}$$

where $\omega = \omega_r + 2\Omega\cos\theta$ is the absolute field of the vortex, $\Omega$ is the angular velocity of rotation of the sphere (for the Earth, $\Omega \approx 7.3\times10^{-5}\sec^{-1}$), $\theta$ is the co-latitude, $\varphi$ is the longitude, $V_\theta = R\frac{d\theta}{dt}, V_\varphi = R\sin\theta\frac{d\varphi}{dt}; \omega_r = \frac{1}{R\sin\theta}\left(\frac{\partial V_\varphi \sin\theta}{\partial\theta} - \frac{\partial V_\theta}{\partial\varphi}\right) = -\Delta\psi$ is the radial component of the local vortical field on the sphere, $\Delta \equiv \frac{1}{R^2}\left(\frac{1}{\sin\theta}\frac{\partial}{\partial\theta}\sin\theta\frac{\partial}{\partial\theta} + \frac{1}{\sin^2\theta}\frac{\partial^2}{\partial\varphi^2}\right)$ is the Beltrami–Laplace operator, and $\psi$ is the stream function (see also Eq. (4)). Equation (1) describes the two-dimensional (2D) motion of a thin spherical layer of fluid.[5] According to Eq. (1), each Lagrangian particle preserves the value of absolute vorticity in a thin layer of liquid on a rotating sphere. The velocity in the spherical reference system $r;\theta;\varphi$ and the components of the vortex can be represented as follows by using Helmholtz's theorem regarding the decomposition of velocity on the potential:[3-5]

$$V_\varphi = -\frac{1}{R}\frac{\partial\psi}{\partial\theta} + \frac{1}{R\sin\theta}\frac{\partial\Phi}{\partial\varphi};$$
$$V_\theta = \frac{1}{R\sin\theta}\frac{\partial\psi}{\partial\varphi} + \frac{1}{R}\frac{\partial\Phi}{\partial\theta} \qquad (2)$$

New exact solutions of the non-linear Eq. (1) are needed to describe geophysical flows that are suitable for testing the resolution of the numerical methods and explaining the physical mechanisms represented by them (see Ref. 44 and the references therein). For the simple case $H = H_0 = const$, the most general known solution of Eq. (1) has the following Ertel form: [44]

$$\psi(t;\theta;\varphi) \equiv \psi_n = \omega_0 R^2 \cos\theta + \sum_{m=-n}^{n} C_{nm} P_n^m(\cos\theta)\exp\left(im\left(\varphi - \sigma_n t\right)\right); \qquad (3)$$

$$P_n^m(z) = \frac{\left(1-z^2\right)^{\frac{m}{2}}}{2^n n!}\frac{d^{n+m}\left(z^2-1\right)^n}{dz^{n+m}}; -1 < z = \cos\theta < 1; n = 1,2,..; m = 0,\pm1,..,\pm n$$
$$\sigma_n = \omega_0 - \frac{2(\omega_0+\Omega)}{n(n+1)}; C_{nm} = C_{nm0} = const \qquad (4)$$

A linear superposition of the associated Legendre–Ferrers functions $P_n^m$, of degree *n* and order *m* of the first kind is used in Eqs. (3) and (4).

The first term in Eq. (3) models the solid-state rotation of a fluid with a constant frequency $\omega_0$. Such rotation can be observed in the zonal transfer in the middle latitudes from west to east in case of positive values of frequency $\omega_0 > 0$.

Note that when $\omega_0 = -\Omega$, corresponding to flow in the reverse direction from east to west, the phase velocity $\sigma_n$ in Eq. (4) is the same for all modes, and does not depend on the number of modes $n$.

For any integer *n*, each of the modes of the wave $\psi_n$ defined in Eq. (3) is an exact solution to the non-linear Eq. (1). However, a linear superposition of such modes for different values of *n* is already not the general solution to Eq. (1), even in case the thin spherical layer of fluid considered in Ref. 44 has a constant depth. The authors of that study proposed seeking the exact solution to Eq. (1) for $H = H_0 = const$ in the form of $\psi = \sum_{n=1}^{\infty} \tilde{\psi}_n = \sum_{n=1}^{\infty} \sum_{m=-n}^{n} C_{nm}(t) P_n^m(\cos\theta) \exp(im\varphi)$, when the time dependent coefficients $C_{nm}(t)$ satisfy the infinite open system of ODEs. Even in case of the simplest superposition $\psi = \tilde{\psi}_1 + \tilde{\psi}_n$ (see Eq. (7) in Ref. 44), only an approximate solution is obtained when it is assumed that the three coefficients $C_{1m}(t)$ of the decomposition of the function $\tilde{\psi}_1$ are known.

We show in Appendix A that contrary to the approach used in Ref. 44, it is possible to obtain the explicit form of a general weak solution, Eqs. (A.1)–(A.7), to Eq. (1) by using point vortices when the dependence of the depth of the layer $H = H_0 h(\theta;\varphi;t)$ on the coordinates is permitted.

According to the solution to Eq. (1) provided by Eqs. (A.1)–(A.7), the intensity and circulation of the vortex increase when the thickness of the layer decreases. Observations have shown an increase in the intensity and circulation of the cyclonic vortex with a reduction in pressure at its center and a corresponding reduction in the level of the isobaric surface above sea level. Note that the representation of the stream function $\psi_R$ given in Eqs. (A.4)–(A7), characterizing the regular vortex field associated with the rotation of the spherical layer, generalizes the exact solution obtained in Ref. 19. The authors of that study used the simpler representation $\psi_R = \psi_0 = -\Omega R^2 \cos\theta$ instead of the common solution provided by Eq. (A.4) to $-\Delta\psi_R + 2\Omega\cos\theta = 0$ when $C_R = 0; h = 1; H = H_0 = const$. Thus, the solution obtained in Ref. 19 does not consider variations in the thickness of the spherical layer at different points on the surface of the sphere over time. The function used in Ref. 19 coincides with the first term of the series in Eq. (A.4) at $n = 1$, when the equalities $A_1 = -\Omega; P_1 = \cos\theta$ are satisfied and $A_n^{(m)} = B_n^{(m)} = 0$ for any $n;m$ in Eq. (A.4). Therefore, even when the thickness of the layer is constant, a more general solution than that in Ref. 19 has the form $\psi_R = -\Omega R^2 (\cos\theta + \sum_{n=2}^{\infty} \tilde{A}_n P_n(\cos\theta)); \tilde{A}_n = \frac{(2n+1)}{2n(n+1)} \int_{-1}^{1} dzz P_n(z)$.

For the sake of simplicity, we consider only the solution of the form[19] $\psi_R = \psi_0 = -\Omega R^2 \cos\theta$, and consider the helical nature of point vortices on the rotating sphere.

For example, when N=1 in Eq. (A.3), the stream function $\psi$ of the single anti-podal point vortex located on the poles of a rotating sphere yields an exact solution to the hydrodynamic equations (see Ref. 5 or Eqs. (B.1) – (B.3) in Appendix B) when $r \to R$: [19]

$$\psi = -\Omega r^2 \cos\theta + \frac{\Gamma_1}{2\pi} \ln\frac{1+\cos\theta}{1-\cos\theta} \tag{5}$$

The first term on the right-hand side (RHS) of Eq. (5) exactly accounts for the effect of rotation of the sphere at frequency $\Omega$.

To consider the helical nature of the point vortex in Eq. (5), we use the representation in Eq. (2). Reference 45 has shown that an exact solution to the same hydrodynamic equations (see Eqs. (B.1)–(B.3) in Appendix B) also corresponds to the velocity field $V_\varphi = 0; V_\theta = \mu_1 / r\pi \sin\theta$ that is created by the point source and sink placed on opposite poles. If $\mu_1 > 0$, the source is located on the north - pole with $\theta = 0$. The potential of the velocity field created by such an anti-podal point source–sink pair on a sphere is as follows:

$$\Phi = -\frac{\mu_1}{2\pi} \ln\frac{1+\cos\theta}{1-\cos\theta} \tag{6}$$

The velocity field obtained from Eq. (2) while considering Eqs. (5) and (6) is as follows:

$$V_\varphi = -\Omega r \sin\theta + \frac{\Gamma_1}{r\pi \sin\theta} \tag{7}$$

$$V_\theta = \frac{\mu_1}{r\pi \sin\theta} \tag{8}$$

Equations (7) and (8) give an exact solution to the hydrodynamic Eqs. (B.1)–(B.3) (see Appendix B). Note that the fact regarding the velocity field in Eq. (7) was used in Ref. 19, and was applied with respect to the field in Eq. (8) in Ref. 45.

We consider only the case of an exact solution to Eq. (1), Eqs. (A.1)–(A.7), for the stream function in the following form:

$$\psi = -\Omega R^2 \cos\theta + \frac{1}{2\pi} \sum_{i=1}^{N} \Gamma_i \ln\frac{1+\cos u_i}{1-\cos u_i} \tag{9}$$

$$\cos u_i \equiv \cos\theta\cos\theta_i + \sin\theta\sin\theta_i\cos(\varphi-\varphi_i)$$

Note that in the case of $N=1$, the stream function in Eq. (9) coincides with the stream function in Eq. (5) in the limit $r \to R$.

Similarly, the potential of the velocity field in Eq. (6) is generalized here for the case of N anti-podal point sources–sinks as follows (when we must make the replacement $\cos\theta \to \cos u_i$ in Eq. (6)):

$$\Phi = -\frac{1}{2\pi}\sum_{i=1}^{N}\mu_i \ln\frac{1+\cos u_i}{1-\cos u_i} \qquad (10)$$

The potential in Eq. (10) is a generalization to the spherical case of the potentials of sources and sinks on the plane, and has been considered in Refs. 6and7. As in Eq. (10), the case $\mu_i > 0$ corresponds to the source in the north on the point with the same spherical coordinates $(\theta_i;\varphi_i)$ as the corresponding point vortex. Thus, in Eqs. (9) and (10), $\theta_i, \varphi_i$ are the spherical coordinates of the anti-podal point HV that are time-dependent functions.

The presence of the point source or sink at a position that coincides with the coordinates of the point vortex does not affect the vorticity, which is still represented by Eq. (A.1) when $C_R = 0; h = 1; \Gamma_i = const, i = 1,2,.., N$. The vortex field is zero for the velocity field described by the potential in Eq. (10) for the point sources and sinks. This is why the consideration of the velocity field created by the potential in Eq. (10) is reduced to the appearance of the additional velocity responsible for the transfer of the absolute vortex field according to Eq. (2), complementing the velocity field created by the stream function in Eq. (9).

To obtain an exact weak solution for Eq. (1) corresponding to the distribution of the vortex field in Eq. (A.1), it is necessary to substitute the distribution of the latter equation into the former, multiply the resulting expression by an arbitrary smooth finite function, and integrate over the spherical surface by considering Eqs. (9) and (10) (see also Ref. 19 and the references therein).

As a result, the 2N coordinates $(\theta_i(t);\varphi_i(t)); i=1,2,..,N$ of all HVs are defined as the solutions to the *2N*-dimensional Hamiltonian system of ordinary differential equations, Eq. (B. 4) (see Appendix B).

## III. Motion of a single HV near the boundary

The problem of considering impenetrable boundaries when modeling the dynamics of point spiral vortices in the thin layer of an ideal incompressible fluid on a rotating sphere is important for using this model to describe and predict geophysical vortical dynamics in the atmosphere and the ocean. An impenetrable boundary is modeled as a line or a set of lines that bound a single-connected region or multi-connected regions on the surface of the sphere, within the framework of the 2D formulation of the problem of vortical dynamics described by only two variables representing the spherical angles. Due to the ideality of the medium, we consider the boundary condition at the specified boundary that is set, for example, by a set of spherical coordinates $\theta=\theta_0(\alpha);\varphi=\varphi_0(\alpha)$ (in this case, the parameter $\alpha\in[0;1]$ characterizes the position of a point on the curve of the boundary). This is in the form of the assumption that the projection of the velocity vector $\vec{V}=(V_\theta;V_\varphi)$ of the particles of the fluid, defined in Eq. (2), on the unit vector of the normal $\vec{n}_0=(n_{0\theta};n_{0\varphi})$ to such a boundary vanishes, $\left(\vec{V}\vec{n}_0\right)_{\theta=\theta_0;\varphi=\varphi_0}=0$.

Appendix A presents a general solution to the AVCE in Eq. (1) by considering the influence of a solid impenetrable boundary on the dynamics of an arbitrary number of point spiral vortices in a thin layer of a fluid of variable thickness on a rotating sphere. The boundary condition used in this case is represented by Eq. (A.10). Equation (A.11) represents the reversal of all components of the velocity of the liquid particles along the boundary to zero, where this corresponds to the possibility of considering the effects of a deviation in the liquid from ideality precisely at the solid boundary of the fluid. However, for the sake of simplicity, we limit ourselves here to considering only the case of the above boundary condition corresponding to the impossibility of fluid flow through the boundary. This allows us to use, as in Ref. 18, the classical method of mirror images for point spiral vortices.

Moreover, we consider only the dynamics of a single HV, and only for cases in which the boundary passes along the Equator of the sphere, a line of arbitrary latitude, or along the meridian line of the sphere. This makes it possible to use the classical method of mirror images to provide a relatively simple account of the above boundary condition.

### A. **Boundary on the Equator**

Let us consider the solution to Eq. (B.4) for the case of a single HV near the solid boundary, placed for the sake of simplicity in the Equatorial plane of the sphere defined by $\theta = \theta_0 = \pi/2; \varphi = \varphi_0 \in [0, 2\pi]$. Here and below, the no-normal-flow boundary condition is considered to describe the influence of impenetrable boundaries on the model. In the specific case when the boundary is located along the Equator, the component of velocity $V_\theta$ of any fluid particle along the meridian must be zero on the boundary. To satisfy the condition of the absence of flow through the boundary, we use the method of mirror reflections. This method introduces mirror symmetric, point vortices, the circulation signs of which are opposite to those of the original vortex, but with the same intensity as for the original vortex. The value and sign of the point sink (or source) for the mirror sink (or source) must match those of the original sink (or source). This method allows us to investigate the interactions between helical vortices and their mirror images.

We define the intensity of the sink and the cyclonic circulation of the point HV as well as its spherical coordinates as $\mu_1 = -\mu < 0; \Gamma_1 = \Gamma > 0; \theta_1 = \theta; \varphi_1 = \varphi$ when, as usual, the frequency of rotation of the sphere is positive by definition, $\Omega > 0$. The values $\mu_2 = -\mu < 0; \Gamma_2 = -\Gamma < 0; \theta_2 = \pi - \theta; \varphi_2 = \varphi$ correspond to the mirror image with respect to the solid boundary when the intensity of the sink of the mirror image of the HV has the same value as that of the HV.[18] In this case, Eq. (B.4) can be rewritten as follows:

$$\begin{aligned} &\frac{d\theta}{d\tau} = \pm \frac{1}{\sin 2\theta}; \\ &\tau = \frac{t\mu}{\pi R^2}; \Omega_1 = \frac{\pi R^2 \Omega}{\mu}; \tilde{\Gamma} = \frac{\Gamma}{\mu} \\ &\frac{d\varphi}{d\tau} = -\Omega_1 + \frac{\tilde{\Gamma}}{\sin\theta \sin 2\theta} \end{aligned} \qquad (11)$$

In case the sign of the first of the above equations is positive, Eq. (11) describes the motion of the HV with cyclonic circulation $\Gamma_1 = \Gamma > 0$ that is conjugated with the radial motion defined by the point sink with intensity $\mu_1 = -\mu < 0$. In case the sign of the first equation above is negative, Eq. (11) describes a combination of the point vortex with cyclonic circulation, but with intensity $\mu_1 = \mu > 0$.

If it is necessary to consider the anti-cyclonic circulation conjugated with the point source, we use the first of the above equations (Eq. (11)) with a negative sign and apply the substitution $\tilde{\Gamma} \to -\tilde{\Gamma}$ to the second.

Fig. 1 shows examples corresponding to the solution of the system of equations, Eq. (11) that can be used to model the observed appearance of the point of inflection on the trajectory of the TC. It is characterized by radial motion of the sink type, and corresponds to sudden changes in trajectory or recurvature.

Fig. 2 shows examples of trajectories corresponding to the above two modifications of the system in Eq. (11), the first of which corresponds to the combination of the point vortex with cyclonic circulation with a point source (denoted by red color). The second modification of the system in Eq. (11) is represented by examples (blue) of anti-cyclonic dynamics that are usually conjugated with radial flow from the center of the vortex that is modeled by a point source.

The solution to Eq. (11) and its modifications for a combination of the point vortex with the point source or sink is as follows:

$$\theta = \frac{1}{2}\arccos\left(\pm 2\tau + \cos 2\theta(0)\right);$$
$$\varphi = \varphi(0) - \Omega_1\tau + \frac{\tilde{\Gamma}}{2}\ln\left|\frac{\left(\sqrt{1 \pm 2\tau + \cos 2\theta(0)} - \sqrt{2}\right)\left(\sqrt{1 + \cos 2\theta(0)} + \sqrt{2}\right)}{\left(\sqrt{1 \pm 2\tau + \cos 2\theta(0)} + \sqrt{2}\right)\left(\sqrt{1 + \cos 2\theta(0)} - \sqrt{2}\right)}\right| \quad (12)$$

The positive sign in Eq. (12) corresponds to the point source and the negative sign to the point sink. $\tilde{\Gamma} > 0$ for the cyclonic point vortex in Eq. (12). The substitution $\tilde{\Gamma} \to -\tilde{\Gamma} < 0$ is needed in Eqs. (11) and (12) in case of an anti-cyclonic vortex.

It follows from Eq. (12) that only in case of a negative sign (or an HV with a point sink) does the HV reach the solid boundary in finite time for $\theta \to \pi/2$ and collapse with its mirror image. This collapse occurs in finite time:

$$\tau = \tau_C = \frac{1}{2}\left(1 \mp \cos(2\theta(0))\right) \tag{13}$$

It also follows from Eq. (12) that only in the case of a negative sign in it and a positive sign in Eq. (13) does the HV reach the solid boundary in finite time for $\theta \to \pi/2$ and collapse with its mirror image.

Fig. 1 shows examples of the trajectories of a spiral vortex with two variants of the initial coordinates. Here the longitudes of both are the same, and they differ only in terms of their initial latitudes. Two other cases are considered, corresponding to different relations between the intensities of the vortex and the sink of the components of a helical vortex. Figs. 1 and 2 show the view from the North Pole so that the Equator in both is represented by a circle, with the North Pole at the center. The horizontal axis in Fig. 1 represents $\varphi = 0$ and the vertical axis represents $\varphi = \pi/2$ (moving eastward when $d\varphi/d\tau > 0$ and westward when $d\varphi/d\tau < 0$).

Figs. 1 and 2 use a scale such that $\theta = \pi/2$ (the Equator) is represented by circle with a radius of one and $\theta = 0$ (North Pole) is at its center.

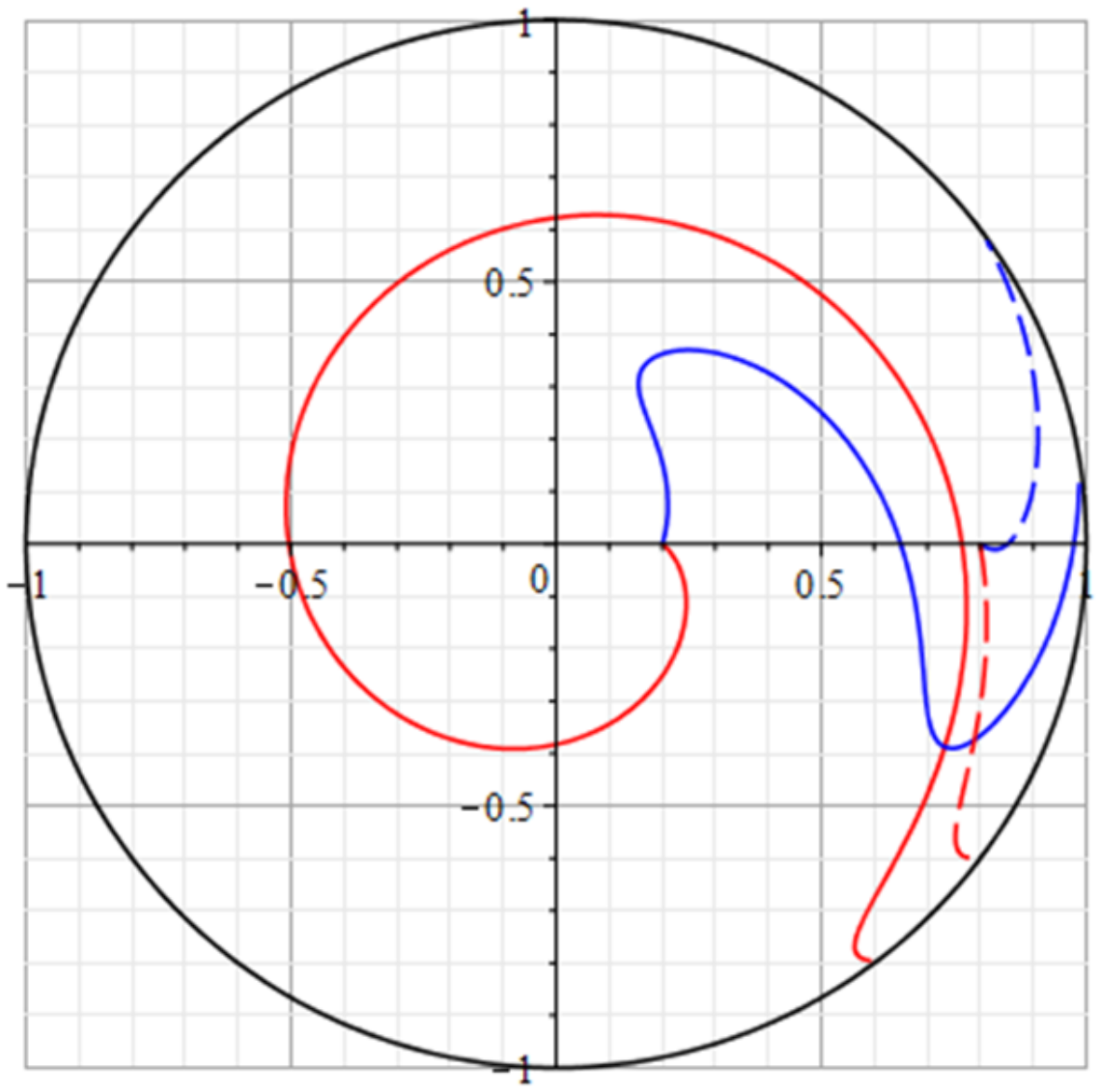

Fig. 1. Tracks of the point helical vortex according to Eq. (11) with a positive sign with initial conditions $\varphi(0) = 0; \theta(0) = 0.1\pi$ (solid line) and $\varphi(0) = 0; \theta(0) = 0.4\pi$ (dashed line). The times

of collapse into the boundary are $\tau_c \approx 0.9$ (solid line) and $\tau_c \approx 0.095$ (dashed line). The cases with $\Omega_1 = 10$, and $\tilde{\Gamma} = 1$ (red) and $\tilde{\Gamma} = 5$ (blue) are considered.

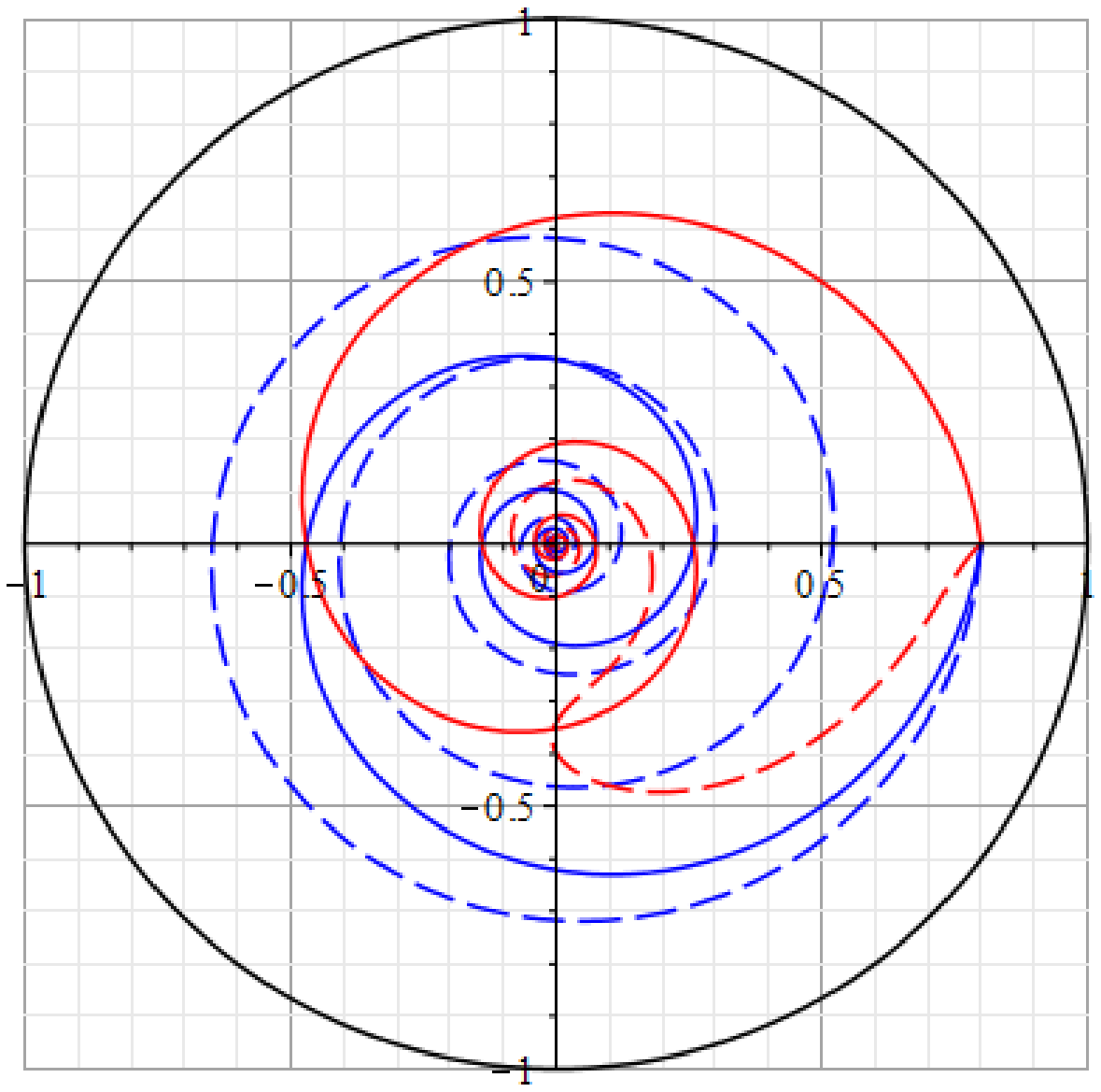

Fig. 2. Solutions of Eq. (11) for combinations of the cyclonic point vortex with the point source (red trajectories) and the anti-cyclonic point vortex with the point source (blue trajectories) under the initial conditions $\varphi(0) = 0; \theta(0) = 2\pi/5, \tau_c \approx 0.9$. The trajectories without the rotation of the sphere, $\Omega_1 = 0$, are represented by the solid lines, and those with a finite frequency of rotation of the sphere, $\Omega_1 = 10$, are represented by the dotted lines.

In the limit $0 < \tau << 1$, the asymptotic representation of Eq. (12) is obtained:

$$\varphi(\tau) = \varphi(0) - \tau\left(\Omega_1 \pm \frac{\tilde{\Gamma}\sqrt{2}}{\sqrt{(1+\cos(2\theta(0)))(1-\cos(2\theta(0)))}}\right) + O(\tau^2) \qquad (14)$$

In Eq. (14), the positive sign corresponds to a combination of the point cyclonic vortex with the point source and the negative sign corresponds to its combination with the point sink.

It follows from Eq. (14) that rotation in the anti-cyclonic direction with $d\varphi/d\tau < 0$ is possible only at the super-threshold of the frequency of rotation of the sphere. The corresponding condition for the violation of chiral symmetry, which determines only the anti-cyclonic direction of the movement of a spiral cyclonic vortex, then has the following form:

$$\Omega_1 > \Omega_{1th} = \frac{\tilde{\Gamma}}{\sqrt{2(1+\cos(2\theta(0)))(1-\cos(2\theta(0)))}} \geq \Omega_{1MINth} = \frac{3\sqrt{3}\tilde{\Gamma}}{8} \qquad (15)$$

In Eq. (15) the minimum value of the threshold of speed of the rotation of the sphere $\Omega_{1MINth}$ is obtained for the initial conditions $\cos\theta(0) = 1/\sqrt{3}; \theta(0) = 54.7^0$. It follows that a violation of chiral symmetry should be expected for the first time under $\Omega_1 > \Omega_{1MINth} = \frac{3\sqrt{3}\tilde{\Gamma}}{8}$, or in dimensional form at a sufficiently high rotational frequency of the sphere in comparison with the rotational frequency of the helical cyclone vortex $\Omega > \omega_{\min th} = 3\sqrt{3}\Gamma / 8\pi R^2$.

References 24 and 25 examined the analogous effects of the realization of dissipative–centrifugal instability (DCI), and provided the primary mechanism for the generation of anti-cyclonic vorticity on quickly rotating planets. The condition for DCI in Ref. 24 is $\Omega > \omega_0$, where $\omega_0$ is the natural frequency of a linear 2D vibrator (modeling the solid-state rotation of a fluid) in a coordinate system rotating at speed $\Omega$.

The tendency of displacement of the point HV with changes in the longitude $\varphi$ in case of a negative value of $d\varphi/d\tau < 0$ in Eq. (14) corresponds to anti-cyclonic rotation. It is clear from Fig. 1 that under the condition imposed by Eq. (15), the initial motion of the point HV corresponds to anti-cyclonic rotation, i.e., clockwise in the Northern Hemisphere, corresponding to the trajectory represented by the solid red line. When Eq. (15) does not hold, the motion is cyclonic in its initial stage of evolution, i.e., counter-clockwise, and is represented in Fig. 1 by the solid blue line. Moreover, Fig. 1 shows that in case of stronger circulation, $\tilde{\Gamma} = 5$, the change in the direction of motion is sharper than under weaker circulation, $\tilde{\Gamma} = 1$.

Fig. 1 also shows that abrupt changes in the direction of motion of the HV substantially depend on its initial position. More precisely, they depend on the distance between it and the boundary. Thus, for $\tilde{\Gamma} = 1$, a sharp change in the direction of motion occurs soon after it is initiated, while changes in its motion occur only in the proximity of the boundary at $\tilde{\Gamma} = 5$.

The mechanism of changes in its direction of motion is related to the symmetry of the HV, defined as the ratio of the intensity of the vortex to that of the sink $\tilde{\Gamma}$, and its initial position on the rotating sphere. This mechanism is unrelated to the character of the background circulation[29] or the vertical structure of the TC.[46]

Note that the structure of the TC is usually defined by the sink at its bottom in the limit of the height of the boundary layer of the atmosphere, and by its source of compensation in the higher layers of the troposphere and the lower layers of the stratosphere.[8, 25] The trajectories of cyclonic and anti-cyclonic HVs interacting with the solid boundary (Equator) are shown in Fig. 2 when their radial flow is modeled by a point source.

The trajectories of the TC directed from the Equator to high latitudes become dominant according to the observation data shown in Fig. 2. Thus, it is natural to model the dynamics of the TC by assuming that the Equator is a solid boundary. A similar model can be used when describing the effects of mutual repulsion between TCs when the distance between their centers is sufficiently large, and point sources are used to represent the interactions between the HVs.

The typical track of a TC in the Northern Hemisphere is an initial north-to-west movement followed by an eventual shift toward the east. Such a shift is referred to as recurvature, and is often explained by changes in environmental flow that steer the TC.[29-31] Reference 46 has shown that even in the absence of background flow, a TC initiated at a sufficiently high latitude can recurve by itself. Differential horizontal advections of planetary vorticity due to the circulation of the TC at different heights lead to the development of vertical wind shear, an anti-cyclone in the upper troposphere, and the asymmetric distribution of convection. The flow associated with the upper tropospheric anti-cyclone on the Equator-ward side of the TC and diabatic heating associated with the asymmetric convection combine to cause the recurvature of the TC. Such knowledge of the intrinsic curvature of the TC is important for forecasting its track when the environmental flow is weak.[46]

The above example proves that there exists a cause of the recurvature of the TC in addition to the ones known. It is conditioned by the distance between the TC and the coastline, and by the ratio of the intensities of the medium rotational and radial motions induced by the TC.

### B. **Boundary on an arbitrary latitude**

We now consider a more general case in which the solid boundary is defined by any value of the complement to the latitude $\theta = 0 < \theta_0 < \pi / 2$. To apply the method of mirror reflections to this scenario, we limit ourselves only to cases in which the spiral vortex and its mirror image, in the initial state and in case of subsequent evolutions, are at a distance from the boundary that is much smaller than the radius of curvature of the circle corresponding to the latitude at which the boundary is located. In contrast to the case discussed above, when the border is located along the Equator (or along the meridian line), only in the limit of a small distance from a boundary located at an arbitrary latitude can we consider the same consideration for the arrangement of the spiral vortex and its mirror image.

Thus, only in the initial instant can the following relationships between the coordinates of the image vortical sink $(\theta_2(\tau);\varphi_2(\tau))$ and the actual vortical sink $(\theta_1(\tau);\varphi_1(\tau))$ hold:

$$\begin{aligned} &\theta_2(0) = 2\theta_0 - \theta_1(0); \\ &\varphi_2(0) = \varphi_1(0) \end{aligned} \qquad (16)$$

We define trajectories of the vortical sink and its mirror image to solve Eq. (B.4) for the initial conditions of the vortex sink $\theta_1(0);\varphi_1(0)$. Equation (16) then holds for its mirror image.

It is necessary to solve the system represented by Eq. (B.7) that follows from Eq. (B.4).

Only at the Equator, when $\theta_0 = \pi / 2$, does the symmetry relation in Eq. (16) hold at any instance when Eq. (B.7) coincides with Eq. (11), with a general solution in the form of Eq. (12) for any initial condition of the vortical sink.

For an arbitrary $\theta_0$, when $\theta_0 \neq \pi / 2$, the condition of equality of the longitudes of HV and its mirror image with respect to the solid boundary is not satisfied at any time $\tau > 0$.

An example of numerical solution of the system (B.7) is represented on Fig.3 wherein at each step of integrating (B.7), a new mirror symmetry condition is defined providing equality $\varphi_1 = \varphi_2$ and no-flow across the solid boundary.

Fig. 3 shows trajectories of the point HV interacting with the solid boundary positioned at 30° latitude in the Northern Hemisphere (note that respective co-latitude is $\theta_0 = 60^0$) as well as the observed TC.

More complex considerations need to be applied in more general cases, when the initial HV is far from the boundary (see Appendix A).

### C. Boundary along the meridian

If the solid boundary is along the meridian with the coordinate $\varphi = \varphi_0$, the system in Eq. (B.8) is derived from that in Eq. (B.7). The system in Eq. (B.8) is consistent with the condition of a solid boundary at arbitrary instances, but only when $\Omega_1 = 0$, i.e., the effects of rotation of the sphere can be neglected. In this case, the boundary condition provides the restriction that the component of velocity of the fluid particles normal to the boundary must be zero, $V_\varphi = 0$, on the boundary.

Fig. 3 shows the results of modeling the observed motion of the TC (denoted by “23” at the end of its trajectory) by solving Eq. (B.8) when the boundary is along the meridian. We also assume a special variation in the intensity of the TC over time.

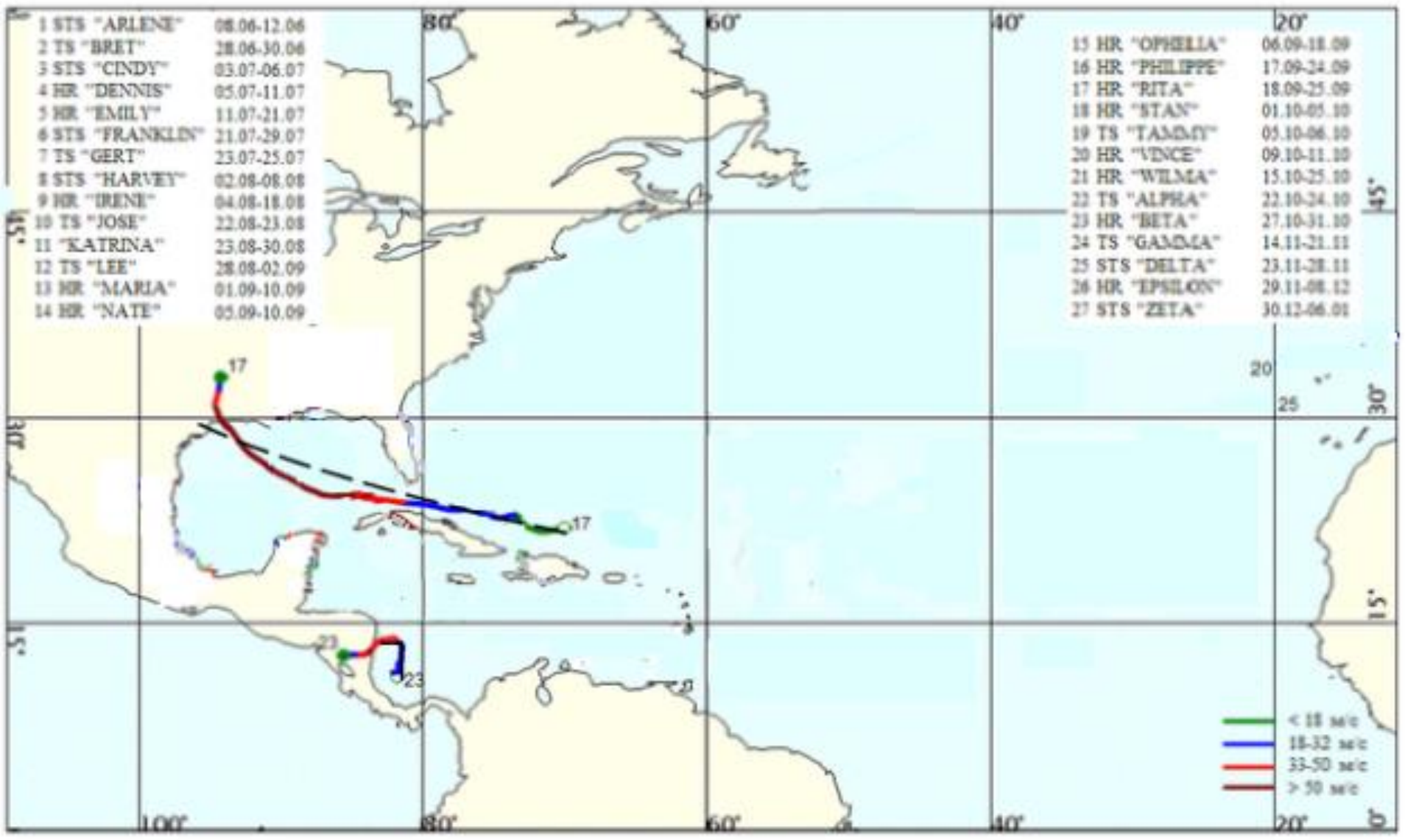

Fig. 3. Modeling the track of cyclone 17 ("RITA") according to solution of Eq. (B.7) with initial data $\theta_1(0) = 68^0; \varphi_1(0) = 70^0 W$ and an HV with parameters $\Omega_1 \equiv \pi R^2 \Omega / \mu = 10; \tilde{\Gamma} \equiv \Gamma / \mu = 1.8$. The modeled track is depicted by the dashed black line when $\tau_c \equiv t\mu / \pi R^2 \approx 0.2$ is the dimensionless time of collapse of the TC into the boundary. The boundary is located along the latitude $90^0 - \theta_0 = 30^0 N$, and the track for the mirror image of the HV is not represented. Observation data for the beginning and end of the track of cyclone 17 are also shown.

We also model the track of cyclone 23 ("BETA") according to the solution to Eq. (B.8) with the initial data $\theta(0) = 78^0; \varphi(0) = 81.6^0 W$ and parameters

$$\Omega_1 = 0, \Gamma(\tau) = \begin{cases} 172.1911529, & \tau \le 0.004 \\ 30, & \tau \le 0.008 \\ 5, & \tau \le 0.012 \\ 0, & \tau > 0.012 \end{cases}.$$

The modeled track is represented by the black line, and $\tau_c \approx 1.9$ is the time of collapse of the TC into the boundary.

The boundary is located along the meridian, and has coordinates $\varphi = \varphi_0 = 83.2^0 W$.

## IV. Interaction of two point helical vortices

We now consider the Fujiwhara effect[1, 9] in greater detail based on the numerical solutions of Eqs. (B.9) and (B.10) (see Appendix B). This effect consists of the interactions between developed TCs if their centers are less than 1700 km apart. For the sake of simplicity, we neglect the effect of the solid boundary on the dynamics of two identical point HVs.

We consider only cases in which only a point sink provides the attraction between HVs independently of their directions of circulation. The HVs have the same parameters of circulation and point sinks. The repulsion between HVs has also been modeled in Ref. 1 by using the point source.

Figs. 4 and 5 show two cases of such interactions between anti-cyclonic vortices with a radial component of the sink when the rotation of the sphere is neglected (Fig. 4) and when it is considered (Fig. 5).

The trajectories of the HVs shown in the figures were obtained by solving Eq. (B.9) with the initial conditions of TCs 17 and 23.

A comparison of Figs. 4 and 5 shows that the rotation of the sphere led to a shift in the center of mass of the system of vortices to the west, and they then fused in finite time.

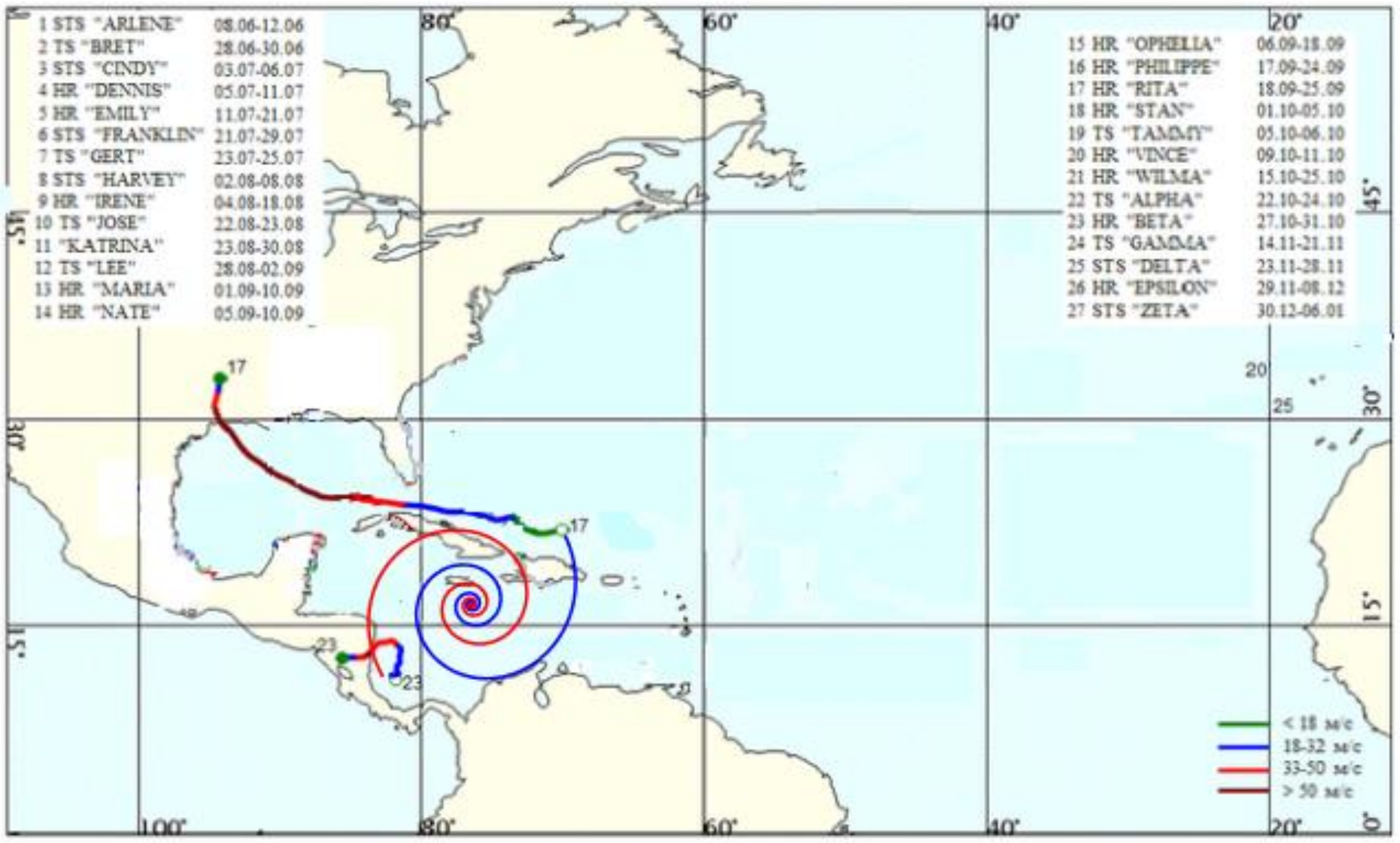

Fig. 4. Interaction between the sinks of point anti-cyclones. We modeled the initial positions of cyclones 17 ("RITA;" blue) and 23 ("BETA;" red) according to Eq. (B.9), with the initial conditions $\theta_1(0) = 68^0, \varphi_1(0) = 70^0 W; \theta_2(0) = 78^0, \varphi_2(0) = 83.2^0 W$ and parameters $\tilde{\Gamma} = 5, \Omega_1 = 0$ when the time of collapse was $\tau_c \approx 0.02$.

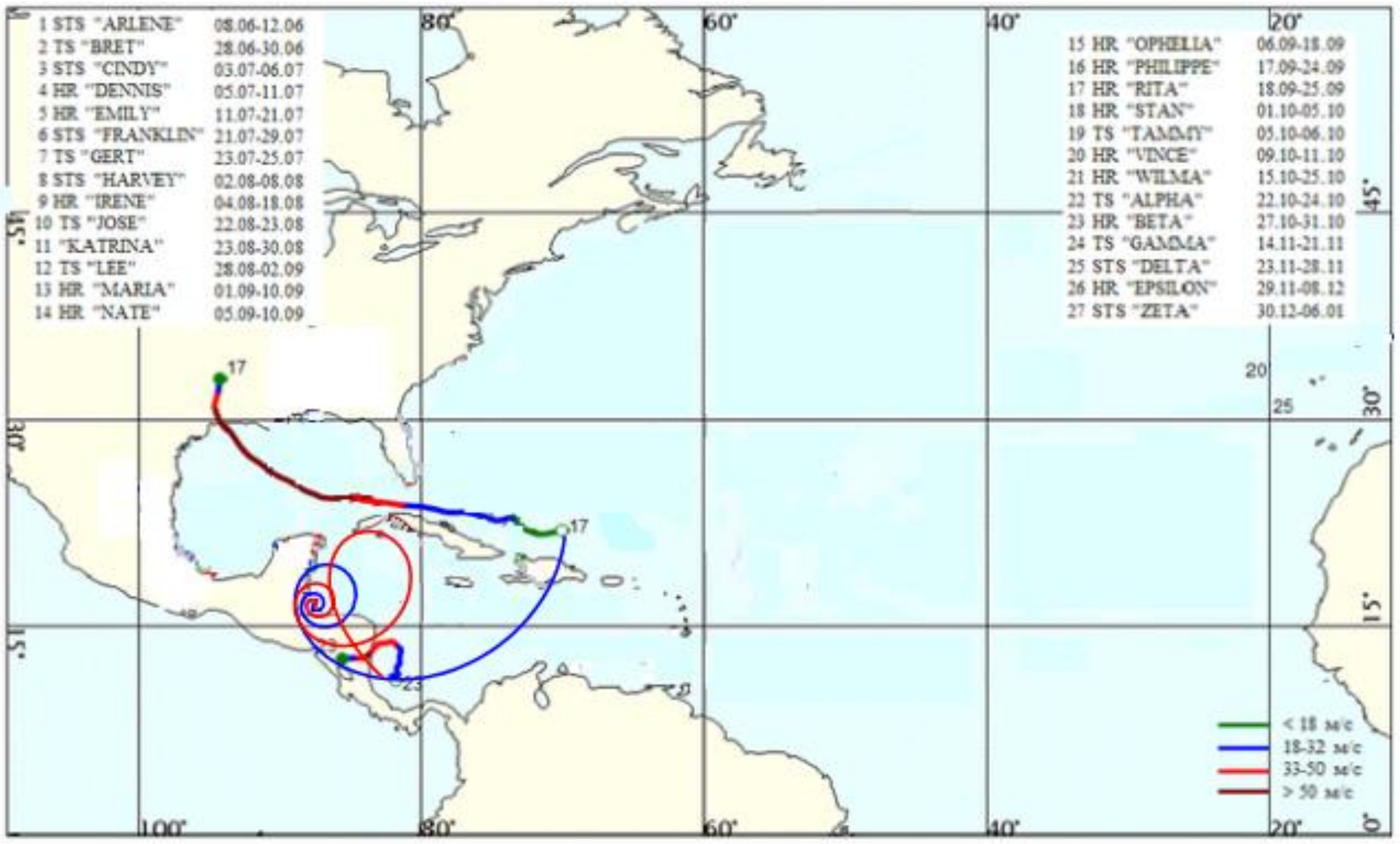

Fig. 5. Interaction between the sinks of point anti-cyclones. We modeled the initial positions of cyclones 17 (“RITA;” blue) and 23 (“BETA;” red) according to Eq. (B.9), with the same initial conditions as in Fig. 4 and the parameters $\tilde{\Gamma} = 5$ and $\Omega_1 = 10$. The collapse time was $\tau_c \approx 0.02$ .

Figs. 6 and 7 show the trajectories of two interacting point HVs of the cyclonic type, obtained by solving the system in Eq. (B.10) and with the initial conditions given in Figs. 4 and 5.

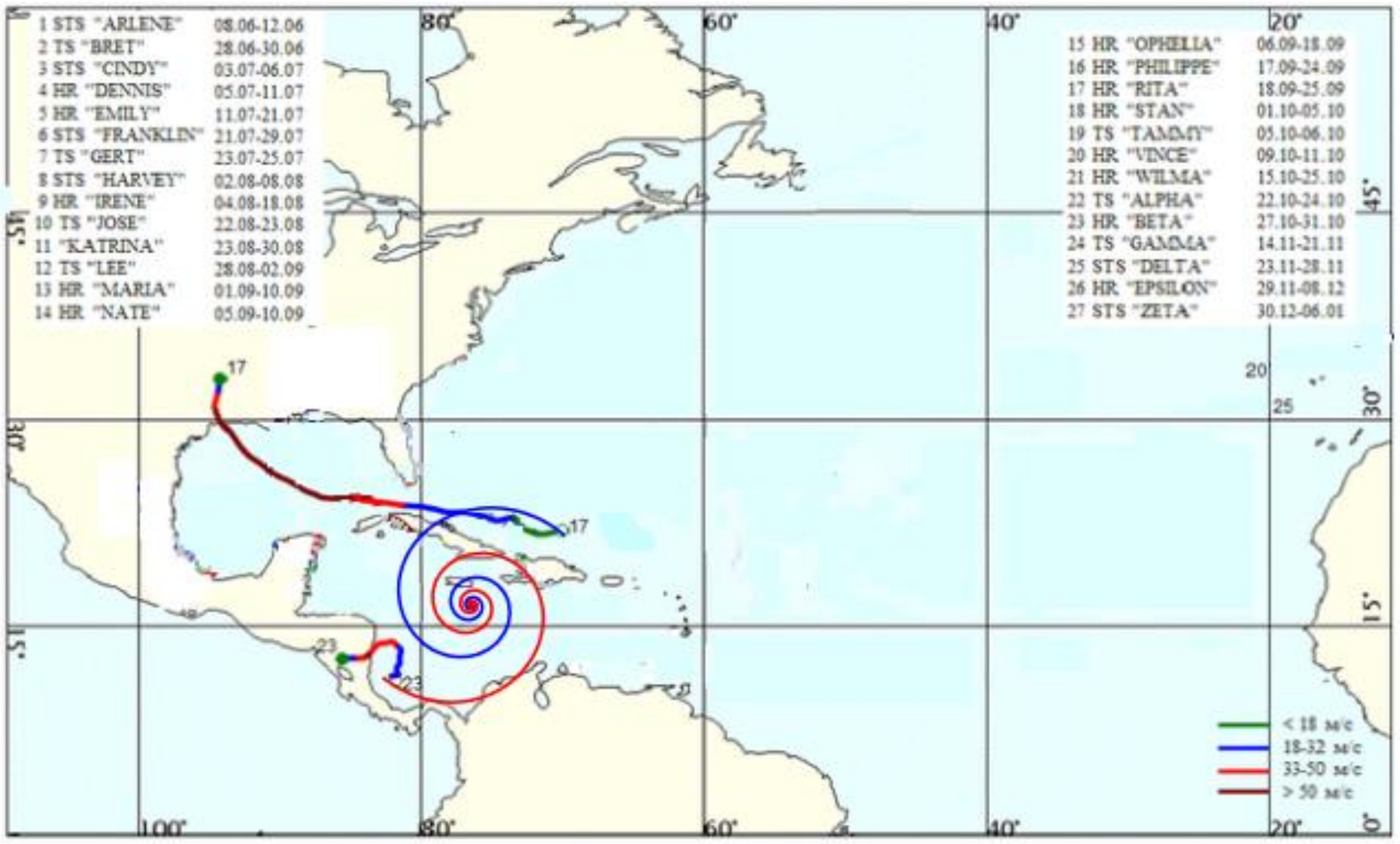

Fig. 6. Interaction between the sinks of point cyclones. We modeled the initial positions of cyclones 17 (“RITA;” blue) and 23 (“BETA;” red) according to Eq. (B.10), with the same initial conditions as in Fig. 4 and the parameters $\tilde{\Gamma} = 5,\ \Omega_1 = 0$, and $\tau_c \approx 0.02$.

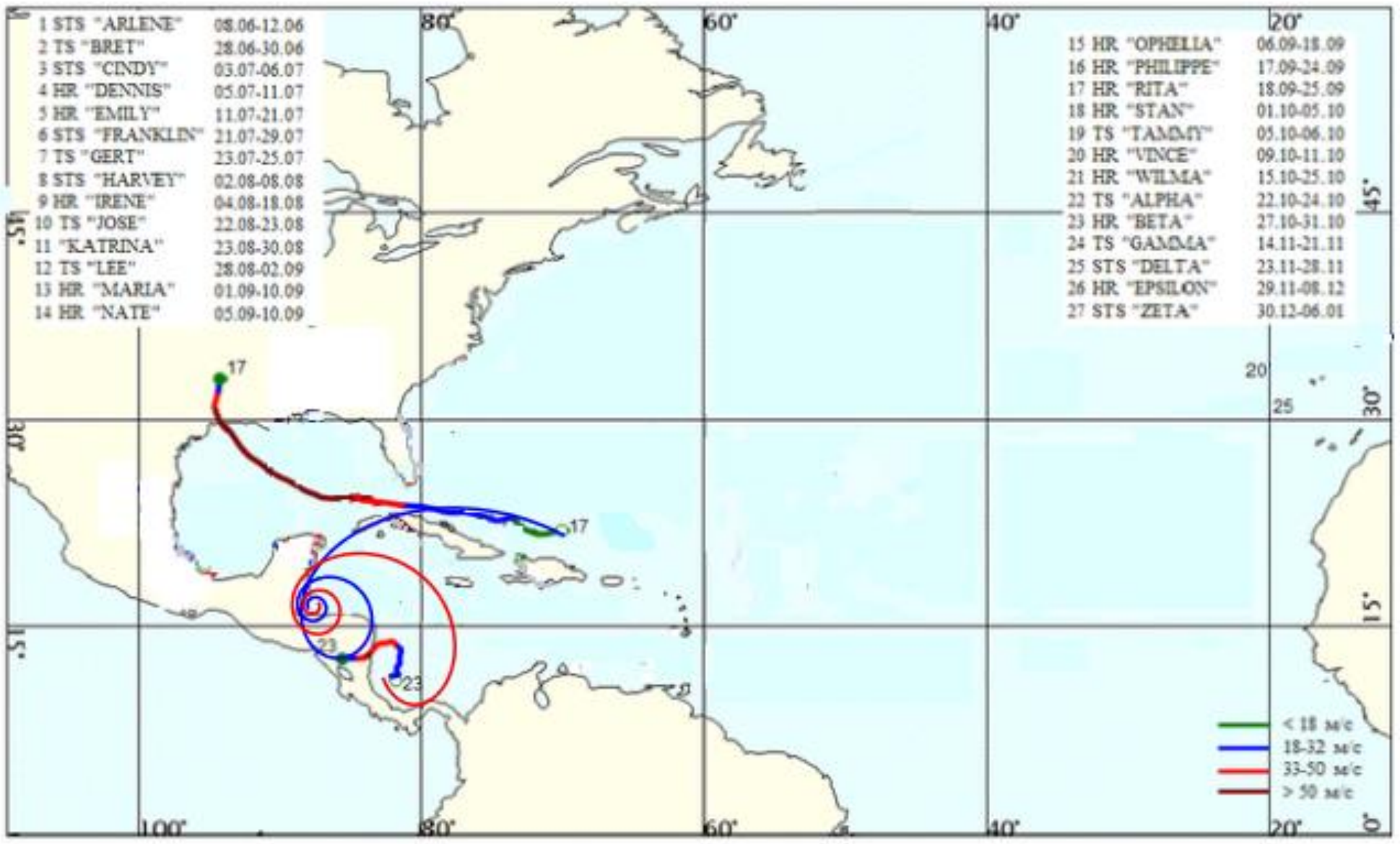

Fig. 7. Interaction between the sinks of point cyclones. We modeled the initial positions of cyclones 17 (HR “RITA;” blue) and 23 (“BETA;” red) according to Eq. (B.10), with the same initial conditions as in Fig. 4 and the parameters $\tilde{\Gamma} = 5, \Omega_1 = 10$, and $\tau_c \approx 0.02$.

Figs. 5 and 7 show that the point of the fusion of two identical HVs progressively shifted westward when accounting for the rotation of the sphere. These results yield the conclusion that the conjugated recurvature of the TC in some cases can be related to its interaction with another TC.

## V. Discussion

The motion of the TC is a complex process, and modeling it requires at least knowledge of the interactions among the circulation of the cyclone, the wind field, Earth’s vorticity field, the underlying surface, and the fields of moist convection. However, Ref. 46 has shown that the short-term motion of the TC can be described by a linear combination of two mechanisms—advection by the environmental wind field and propagation due to interactions with the Earth’s vorticity field—through beta-plane approximation. For example, the beta effect in the Southern

Hemisphere generally causes a west-to-southwest deviation from the basic advection of the current. Deviations in the Northern Hemisphere occur from the west to the northwest, north, and northeast. The match between the representations of the model and the observation data reported in Ref. 46 does not mean that the authors were able to completely determine the resulting trajectory of the TC. We have proposed here a mechanism for the strong, long-range interactions of a TC with the land that provides an alternative for explaining the trajectories of the TC near the coastline.

The results obtained here prove that the long-range interactions of the TC with the mainland and other TCs have an impact on their recurvature and sudden landfall. The interaction of the TC with the mainland has been considered in the literature to date only after it has crossed the coastline.[29-39, 46] The relevant studies have not solved the practically important problem of providing a mechanism for the abrupt changes in its trajectory. This is especially important when the TC suddenly makes landfall, where this can lead to a major disaster. The long-range interactions between the TC and the coastline might explain the mechanism of such changes.

The proposed model of the mechanism of interactions between the TC and the mainland considers the effects of the large-scale circulation of the TC, and relates its recurvature and landfall to the Coriolis force. A TC can be attracted to the boundary of the ocean similarly to how small-scale spiral vortices (tornadoes, in particular[18]) are attracted to a solid and rigid boundary.

In addition to identifying external factors that influence the trajectory of the TC, Ref. 46 (see also Ref. 32) noted that its dynamics depend on its parameters. For example, the maximum wind speed in the field of distribution of the radial velocity of the TC does not have a significant influence on its trajectory under a fixed value of its field of velocity of circulation. At the same time, changes in the latter in conjunction the intensity of the vortex of the TC and its effective size significantly affect its dynamics. Thus, even a modest change in its outer circulation can cause the larger vortex to move westward and poleward more quickly than the reference vortex.[32, 46]

A similar conclusion regarding the significance of the impact of circulation on the dynamics of the TC follows from the exact solution for the dynamics of the HV on the rotating sphere shown in Fig. 1. When the circulation of the HV increases under a fixed velocity of rotation of the sphere and the sink of the HV, its trajectory is characterized by changes in its direction of motion (in Fig. 1, this trajectory is shown in blue). Moreover, deviations in its trajectory from the west to the northwest in the Northern Hemisphere, which are related to the beta effect considered in Ref. 46, are presented in Fig. 2. The displacement toward the north in Fig. 2 does not occur due to the beta effect, but due to the interactions of the TC with the Equator, which acts as an effective solid boundary. We modeled the dynamics of the TC while precisely accounting for the dependence of the Coriolis force on the latitude. This allowed us to determine conditions for the applicability of the conclusions obtained in beta-plane approximation (see Appendix B, where the exact solution to Eq. (1) is obtained using this approximation).

Moreover, the rules defined in Ref. 32 to determine conditions influencing abrupt changes in the trajectory of the TC and those defined in Ref. 33 to determine the possibility of the TC entering a coastline have emphasized the importance of its initial position as characterized by the latitude and longitude. When modeling the dynamics of the TC, we found that its initial position and its distance from the boundary were significant variables. It follows from Fig. 1 that abrupt changes in the trajectory of the TC occur only when its initial position is sufficiently far from the solid boundary and its circulation is sufficiently large.

Frequently observed interactions between TCs may cause complicated storm-related motion, such as loops, and sharp changes in its direction and translatory speed.[1, 9, 42, 47-50] For example, nearly one-third of all TCs in the Western Pacific region in 1984 had a period of interactions with another cyclone.[50] Errors of 1850 km in the determine of the TC position or larger at 72 h might have occurred in these cases (see also the references in Ref. 50). Mutual rotation is usually cyclonic, but cases of the anti-cyclonic rotation of a cyclonic pair have also been observed. References 47–50 showed that considering the vertical structure of the TC yielded interactions between TCs, the effective radius of which increased from 300 km in the barotropic case when the vertical structure was not considered to 1000 km when it was considered. Divergent winds in the baroclinic models were responsible for the mutual attraction between vortex pairs. The interaction between vortices was found to be almost unaffected by the beta effect, and they might have moved as a single system.

The fusion of similar two vortices is explained in the theory of hydrodynamics of an ideal, incompressible, 2D liquid by substituting point vortices with round vortices of a finite size[40] or elliptical vortices[42] with a uniform velocity field. The fusion of vortices is possible only when the initial distance between their centers does not exceed a threshold proportional to their sizes. In particular in laboratory experiments,[1] vortices with the same direction of circulation in water have been found to fuse only when they are sufficiently close to begin with (due to viscosity, the vortices are not assumed to be point vortices). However, the mechanism of attraction cannot be fully explained by using vortices of finite size.[42] Reference 43 showed that the symmetry of the velocity fields of the vortices is crucial for their fusion in addition to their sizes, especially in case of point vortices. The authors showed that two same extremely small elliptical point vortices can fuse for definite combinations of the intensities of their monopole and quadrupole components. This explains the Fujiwhara effect, even in case of two point vortices in an ideal liquid without sources/sinks and without viscosity to yield finite-sized vortices. The authors obtained the conditions for the fusion of as well as the repulsion between two identical point vortex quadrupoles (PVQs). Each such PVQ was an infinitely small elliptic vortex, the measure of spatial localization of which tended to zero.[43] Reference 7 noted a similar case of the mutual attraction or repulsion between identical HVs. However, despite differences, the theories in both Refs. 7 and 43 explain the Fujiwhara effect based on properties of the symmetry of point vortices that are not related to their finite sizes.

In this study, we modeled the observed effects of the fusion of the centers of two TCs, as in Ref. 7, based on the dynamic interactions between identical point HVs. However, contrary to the theories in Refs. 6, 7, and 18, we obtained the expressions for vortical dynamics not from kinematic considerations, but from weak exact solutions of their hydrodynamical equations. Moreover, we considered a sphere instead of a plane to account for the effect of its rotation. For this reason, we generalized the equations of anti-podal point vortices considered in Ref. 19 to the case in which the HVs had tangential as well as radial velocity.

Figs. 4–7 show the results of computations by using the model of two interacting HVs. Figs. 4 and 5 show the trajectories of the HVs with anti-cyclonic circulation that resulted in the clockwise rotation of the overall system for the case considered in the Northern Hemisphere. Figs. 6 and 7 show the motions of HVs with cyclonic circulation in the counter-clockwise direction. Accounting for the rotation of the sphere yielded a shift in the center of symmetry of the vortical system to the west, as shown in Figs. 5 and 7. They show the results of a model that considered the rotation of the sphere.

The forms of the trajectories in Figs. 5 and 7 show that considering the rotation of the sphere not only led to a shift in the system as a whole, but also noticeably complicated the shapes of trajectories of the HVs of the two interacting TCs.

Note that Figs. 5 and 7 do not show shifts of the vortical system in the meridian direction. As in Ref. 19, we considered a variant of the solution to the AVCE for a rotating sphere such that the shift in the arbitrary constant velocity of the HVs in the meridian direction was set to zero. This is why considering the rotation of the sphere resulted in a shift only in the center of the vortical system to the west, as shown in Figs. 5 and 7 for two interacting anti-cyclonic and cyclonic vortices, respectively. By comparison, Figs. 4 and 6 show the interaction of two vortices when the rotation of the sphere was not taken into account.

## VI. Conclusion

The minimalistic, pure dynamical approach proposed here does not pretend to provide a comprehensive explanation of the observation data (details of the landfall of the TC in particular).

The results obtained here help identify the significant role of the rotation of a sphere on the dynamics of the HVs in their interactions with each other and with the boundary. In contrast to those in Ref. 51, the results here are based on an exact solution of the equation of conservation of absolute vorticity on a rotating sphere. The outcomes related to the trajectories of the TCs are interesting from the point of view of their evolution in relation to the mechanism of formation of a large-scale anti-cyclonic vortex, the stability of which has been studied in Refs. 52 and 53. The possibility of self-similar solutions in a dynamical system of HVs on a rotating sphere and its beta-plane approximation requires special consideration, as in studies on the dynamics of a point vortex on a plane and a sphere[53-57] as well as those of HVs on a plane.[7] The dynamic system obtained here and solutions to it can be useful in problems of turbulence, the collapse of vortices, anomalous turbulent energy dissipation,[58-65] and synoptic meteorology.[66]

The first main result of this paper is the description of the motion of the vortex on a sphere with impenetrable boundaries, which is an important problem in vortical dynamics due to its relevance to modeling geophysical flows—especially oceanographic flows. Oceanic eddies frequently interact with such topographies as ridges and coastlines, and such interactions can play an important role in ocean circulation and various other oceanic processes.[67-71] At the same time, unlike the methods used in Ref. 67 and Refs. 69–71 (contour dynamics and stereographic projection), we considered the rotation of the sphere when modeling the dynamics of point spiral vortices. As in Refs. 69–71, we considered the influence of a solid impenetrable boundary by using the classical method of mirror images (see also Ref. 72) when analyzing the influence of the boundary on the dynamics of an isolated spiral vortex. To the best of our knowledge, the presence of a point sink or source in combination with a point vortex has not been considered in the relevant research to date when considering the dynamics of the corresponding point spiral vortex on a rotating sphere in a limited area on the surface of the latter. The possibility noted in Ref. 69, of using the mirror image method to account for the effect of an impenetrable boundary only to describe vortices in the ocean, was expanded here and linked to the problem of describing the interactions of localized and intense atmospheric vortices with the coastline. The very possibility of such an application of the mirror image method to model the dynamics of tropical cyclones arises due to the observed feature of tropical cyclones, whereby their intensity decreases when they reach the boundary of the ocean.

The second important result of this paper was obtained in analytical form. We described the cyclone–anti-cyclone asymmetry (Eq. (15)) of the cyclonical track based on the exact solution (Eq. (12)) describing the evolution of a cyclonical HV (with sink-type radial flow close to it), which arose due to its interactions with the solid boundary located along the Equator.

The above condition is clearly similar to the previously obtained condition for cyclone–anti-cyclone vortical asymmetry observed in the Earth's atmosphere, the ocean, and in the atmospheres of other rather rapidly rotating planets.[24, 25, 52, 53, 73, 74] It is different from the criterion obtained in Ref. 24, for the realization of dissipative–centrifugal instability, for two reasons. First, the condition in Eq. (15) was obtained specifically for point spiral vortices, and not for vortices with a rotating solid core. Second, Eq. (15) considers the effects of sphericity, which is manifested in the presence of a significant dependence of the threshold frequency of rotation of the sphere on the initial position of the disturbance in the vortex modeled by using a point spiral vortex. At the same time, it turns out that the effect of cyclone–anti-cyclone asymmetry is more complicated near the Equator and the poles than in the middle latitudes. Thus, if the disturbance in the vortex is located directly near the Equator or either of the poles, the threshold of the speed of rotation of the sphere increases indefinitely. This offers additional possibilities for analyzing the mechanisms of the observed cyclone–anti-cyclone asymmetry in the atmospheres of the Earth, Jupiter, and Saturn (see Ref. 74).

**Appendix A: General solution of the absolute vortex conservation equation (AVCE)**

1. **Without a boundary.** In the absence of any boundary in the fluid layer on a rotating sphere, we seek a solution to the AVCE (Eq. (1)) by using the following representation of the stream function and the vortex field:

$$\begin{aligned} &\omega = \omega_R + \omega_S; \psi = \psi_R + \psi_S \\ &\omega_S = -\Delta \psi_S = h(\theta;\varphi;t)\hat{L}(\delta) \\ &\omega_R = -\Delta \psi_R + 2\Omega \cos\theta = C_R H_0 h(\theta;\varphi;t); C_R = const; H \equiv H_0 h \end{aligned} \quad (A.1)$$

$$\psi_S = \sum_{i=1}^{N} \Gamma_i h(\theta_i;\varphi_i;t) \ln\left(\frac{1+\cos u_i}{1-\cos u_i}\right); \cos u_i \equiv \cos\theta \cos\theta_i + \sin\theta \sin\theta_i \cos(\varphi - \varphi_i); \quad (A.2)$$

$$\begin{aligned} &h(\theta_i;\varphi_i;t) = h(\pi - \theta_i;\varphi_i + \pi;t) \\ &\Gamma_i(t) h(\theta_i(t);\varphi_i(t);t) = const \end{aligned}$$

$$\hat{L}(\delta) \equiv \frac{1}{R^2} \sum_{i=1}^{N} \frac{\Gamma_i}{\sin\theta_i} \left(\delta(\theta - \theta_i)\delta(\varphi - \varphi_i) - \delta(\theta + \theta_i - \pi)\delta(\varphi - \varphi_i - \pi)\right); \varphi_i \equiv \varphi_i(t); \theta_i \equiv \theta_i(t) \quad (A.3)$$

$$\psi_R = R^2 \sum_{n=1}^{\infty} \left[ A_n P_n(\cos\theta) + \sum_{m=1}^{n} P_n^{(m)}(\cos\theta)\left(A_n^{(m)} \cos m\varphi + B_n^{(m)} \sin m\varphi\right) \right]; P_n \equiv P_n^{(0)} \quad (A.4)$$

$$A_n = \frac{(2n+1)}{4\pi n(n+1)} \int_{-\pi}^{\pi} d\varphi \int_0^{\pi} d\theta \sin\theta P_n(\cos\theta) F(\theta;\varphi); F \equiv C_R H_0 h(\theta;\varphi;t) - 2\Omega \cos\theta; \quad (A.5)$$

$$A_n^{(m)} = \frac{(2n+1)(n-m)!}{2\pi n(n+1)(n+m)!} \int_{-\pi}^{\pi} d\varphi \cos m\varphi \int_0^{\pi} d\theta \sin\theta P_n^{(m)}(\cos\theta) F \quad (A.6)$$

$$B_n^{(m)} = \frac{(2n+1)(n-m)!}{2\pi n(n+1)(n+m)!} \int_{-\pi}^{\pi} d\varphi \sin m\varphi \int_0^{\pi} d\theta \sin\theta P_n^{(m)}(\cos\theta) F \quad (A.7)$$

In Eq. (A.3), $\delta$ is the Dirac delta function. The exact weak solution of Eq. (1) in the form of Eqs. (A.1) – (A.7) is valid only for the case in which the conditions given in Eq. (A.2), regarding the circulation of the vortices $\Gamma_i(t); i = 1,2,..,N$ and the depth of the layer $H = H_0 h(\theta;\varphi;t)$ at points on the sphere $\theta = \theta_i(t); \varphi = \varphi_i(t), i = 1,2,...,N$, are realized. The physical meaning of the first condition specified in Eq. (A.2) corresponds to the assumption that at diametrically anti-podal points of the sphere, the thickness of the medium layer must coincide with each other. The second condition in Eq. (A.2) is the assumption of a correlation between the thickness of a thin spherical layer $H_i = H_0 h(\theta_i(t);\varphi_i(t);t)$ and the magnitude of circulation $\Gamma_i$ of the vortex field at $\theta = \theta_i(t); \varphi = \varphi_i(t), i = 1,2,...,N$ for each point vortex on the sphere.

2. **With a solid boundary.** To describe the dynamics of point spiral vortices, we take into account the solid boundaries and consider a generalization of the solution in Eqs. (A.1)–(A.7). This considers the condition for the disappearance of velocity along the boundary. We seek a solution to Eq. (1) while using the following, new representation of the stream function that is included in the definition of the velocity field (Eq. (2)) instead of the relation in Eq. (A.1) in this case:

$$\psi = \psi_S + \psi_R + \psi_B \tag{A.8}$$

The first two terms on the right side of Eq. (A.8) are defined in Eq. (A.1), and the form of the third term is determined by the solution to the Beltrami–Laplace equation $\Delta \psi_B = 0$.

In general, the stream function $\psi_B$ has the following representation:

$$\psi_B = R^2 \sum_{n=1}^{M} \left[ A_{Bn} P_n(\cos\theta) + \sum_{m=1}^{n} P_n^{(m)}(\cos\theta) \left( A_{Bn}^{(m)} \cos m\varphi + B_{Bn}^{(m)} \sin m\varphi \right) \right] \tag{A.9}$$

The solution of Eq. (A.9) coincides with the representation of Eq. (A.4), but the arbitrary constants included in Eq. (A.9) must already have been determined from the boundary conditions (from the requirement $V_\theta(\theta = \theta_0(\alpha); \varphi = \varphi_0(\alpha); t) = 0; V_\varphi(\theta = \theta_0(\alpha); \varphi = \varphi_0(\alpha); t) = 0$ along the boundary for the velocity field in Eq. (2) and by considering Eq. (A.8)) with the following form:

$$\left(\frac{\partial \psi_B}{\partial \theta}\right)_{\theta=\theta_0(\alpha);\varphi=\varphi_0(\alpha)} = \left[-\frac{\partial(\psi_S+\psi_R)}{\partial \theta} + \frac{1}{\sin\theta}\frac{\partial \Phi}{\partial \varphi}\right]_{\theta=\theta_0(\alpha);\varphi=\varphi_0(\alpha)} \equiv F_{1B} \qquad \text{(A.10)}$$

$$\left(\frac{\partial \psi_B}{\partial \varphi}\right)_{\theta=\theta_0(\alpha);\varphi=\varphi_0(\alpha)} = -\left[\frac{\partial(\psi_S+\psi_R)}{\partial \varphi} + \sin\theta\frac{\partial \Phi}{\partial \theta}\right]_{\theta=\theta_0(\alpha);\varphi=\varphi_0(\alpha)} \equiv F_{2B} \qquad \text{(A.11)}$$

In Eqs. (A.10) and (A.11), the explicit form of the functions $\psi_S;\psi_R;\Phi$ included on the right side is determined from the exact solution to Eq. (1) provided by Eqs. (A.1)–(A.7) in case of the absence of boundaries and a potential (Eq. (10)). In this case, the form of the curve of the boundary is set on the sphere in the parametric form $\theta=\theta_0(\alpha);\varphi=\varphi_0(\alpha);\alpha\in[0,1]$.

Let us consider, for simplicity, the case in Eq. (A.9) when the constant coefficients of expansion in series have the following form:

$$A_{Bn}=0;$$
$$A_{Bn}^{(m)}=A=\frac{(F_{1B}d_4-F_{2B}d_2)}{R^2(d_1d_4+d_2d_3)};$$
$$B_{Bn}^{(m)}=B=\frac{(F_{1B}d_3+F_{2B}d_1)}{R^2(d_1d_4+d_2d_3)};$$
$$d_1=\sum_{n=1}^{M}\sum_{m=1}^{n}\frac{dP_n^{(m)}(\cos\theta_0)}{d\theta_0}\cos m\varphi_0;\, d_2=\sum_{n=1}^{M}\sum_{m=1}^{n}\frac{dP_n^{(m)}(\cos\theta_0)}{d\theta_0}\sin m\varphi_0$$

$$d_3=\sum_{n=1}^{M}\sum_{m=1}^{n} mP_n^{(m)}(\cos\theta_0)\sin m\varphi_0;\, d_4=\sum_{n=1}^{M}\sum_{m=1}^{n} mP_n^{(m)}(\cos\theta_0)\cos m\varphi_0 \qquad \text{(A.12)}$$

In the case under consideration, the exact solution of Eq. (1), which satisfies the boundary conditions in Eqs. (A.10) and (A.11) with an arbitrary shape of the curve of the boundary on the surface of the sphere, is described by the velocity field in Eq. (2), in which the current function is determined from Eq. (A.8) by taking into account the representations in Eq. (A.2), Eqs. (A.4)–(A.7), and Eqs. (A.9) and (A.12).

**Appendix B: Dynamics of point helical vortices on a rotating sphere**

**1. Hydrodynamical equations on a sphere.** Consider a spherical coordinate system $(r,\theta,\varphi)$ rigidly connected to the globe, with the origin at the center of the globe. On the assumption that only the radial component of the velocity field is equal to zero, $V_r = 0, V_\theta \neq 0, V_\varphi \neq 0$ , the equations can be represented in the following form:[5]

$$\frac{V_\theta^2 + \left(V_\varphi + \Omega r sin\theta\right)^2}{r} = \frac{1}{\rho_0}\frac{\partial p}{\partial r} \qquad \text{(B.1)}$$

$$\frac{\partial V_\theta}{\partial t} + \frac{V_\theta}{r}\frac{\partial V_\theta}{\partial \theta} + \frac{\left(V_\varphi + \Omega \mathrm{rsin\theta}\right)}{rsin\theta} \cdot \frac{\partial V_\theta}{\partial \varphi} - \frac{\left(V_\varphi + \Omega \mathrm{rsin\theta}\right)^2 cot\theta}{r} = -\frac{1}{\rho_0 r}\frac{\partial p}{\partial \theta} \qquad \text{(B.2)}$$

$$\frac{\partial V_\varphi}{\partial t} + \frac{V_\theta}{r}\frac{\partial (V_\theta + \Omega \mathrm{rsin\theta})}{\partial \theta} + \frac{V_\varphi + \Omega \mathrm{rsin\theta}}{\mathrm{rsin\theta}}\frac{\partial V_\varphi}{\partial \varphi} + \frac{V_\theta (V_\varphi + \Omega \mathrm{rsin\theta})}{r} cot\theta = -\frac{1}{\rho_0 r sin\theta}\frac{\partial p}{\partial \varphi} \qquad \text{(B.3)}$$

In Eqs. (B.1)–(B.3), $\rho_0 = const$ is density and $\Omega = \mathrm{const}$ is a constant speed of rotation.

**2. Generalization of weak solutions of the hydrodynamics equations obtained in Ref. 19.** The functions $\theta_i(t), \varphi_i(t)$ are defined as the solutions to the following *2N*-dimensional Hamiltonian system of ordinary differential equations:

$$\frac{d\theta_i}{dt} = \frac{1}{\pi R^2}\sum_{\substack{k=1 \\ k\neq m}}^{N} \frac{1}{(1-\cos^2 u_{ik})}\left(\mu_k\left(\cos\theta_k \sin\theta_i - \sin\theta_k \cos\theta_i \cos(\varphi_i - \varphi_k)\right) - \Gamma_k \sin\theta_k \sin(\varphi_i - \varphi_k)\right)$$

$$\frac{d\varphi_i}{dt} = -\Omega - \frac{1}{\pi R^2}\sum_{\substack{k=1 \\ k\neq i}}^{N} \frac{1}{(1-\cos^2 u_{ik})}\left(\Gamma_k\left(\cot\theta_i \sin\theta_k \cos(\varphi_i - \varphi_k) - \cos\theta_k\right) - \mu_k \frac{\sin\theta_k \sin(\varphi_i - \varphi_k)}{\sin\theta_i}\right) \qquad \text{(B.4)}$$

where $\cos u_{ik} = \cos\theta_i \cos\theta_k + \sin\theta_i \sin\theta_k \cos(\varphi_i - \varphi_k)$.

The system in Eq. (B.4) for $\mu_i = 0$ corresponds to the system derived in Ref. 19 for N pairs of APVs. Thus, the dynamical system in Eq. (B.4) gives the generalization of the theory[19] for cases with non-zero sinks or sources, where the parameter $\mu_i \neq 0$ characterizes the intensity of the spiral component of each point spiral vortex.

The system in Eq. (B.4) has invariants corresponding to the invariants of the dynamic system for N APVs with coordinates $\left(x_i(t); y_i(t)\right), i = 1,..., N$ on the plane considered in Ref. 7 as:

$$I_1 = \sum_{i=1}^{N}\left(\mu_i x_i - \Gamma_i y_i\right); I_2 = \sum_{i=1}^{N}\left(\mu_i y_i + \Gamma_i x_i\right) \qquad \text{(B.5)}$$

The invariants in Eq. (B.4) can be represented by Eq. (B.5) provided that Cartesian and spherical coordinates are related by

$$x_i(t) = R\sin\theta_i(t)\cos\varphi_i(t); y_i(t) = R\sin\theta_i(t)\sin\varphi_i(t) \qquad \text{(B.6)}$$

**3. Interaction of two helical vortices.** Consider the system in Eq. (B.4) for the case of $N = 2$. The two point helical vortices have intensities of sinks of $\mu_1 = \mu_2 = -\mu < 0$, and the same value but with opposite circulation signs $\Gamma_1 = -\Gamma_2 = \Gamma > 0$. Then, from Eq. (B.4), the following system of ordinary differential equations of the fourth order follows:

$$\begin{aligned}
\frac{d\theta_1}{d\tau} &= \frac{1}{A}\left(-\cos\theta_2\sin\theta_1 + \sin\theta_2\cos\theta_1\cos(\varphi_1-\varphi_2) + \tilde{\Gamma}\sin\theta_2\sin(\varphi_1-\varphi_2)\right);\\
\frac{d\theta_2}{d\tau} &= \frac{1}{A}\left(-\cos\theta_1\sin\theta_2 + \sin\theta_1\cos\theta_2\cos(\varphi_1-\varphi_2) + \tilde{\Gamma}\sin\theta_1\sin(\varphi_1-\varphi_2)\right);\\
\frac{d\varphi_1}{d\tau} &= -\Omega_1 + \frac{1}{A}\left(\tilde{\Gamma}\left(\cot\theta_1\sin\theta_2\cos(\varphi_1-\varphi_2) - \cos\theta_2\right) - \frac{\sin\theta_2\sin(\varphi_1-\varphi_2)}{\sin\theta_1}\right);\\
\frac{d\varphi_2}{d\tau} &= -\Omega_1 - \frac{1}{A}\left(\tilde{\Gamma}\left(\cot\theta_2\sin\theta_1\cos(\varphi_1-\varphi_2) - \cos\theta_1\right) - \frac{\sin\theta_1\sin(\varphi_1-\varphi_2)}{\sin\theta_2}\right)\\
A &= 1-\cos^2 u_{12}; \cos u_{12} = \cos\theta_1\cos\theta_2 + \sin\theta_1\sin\theta_2\cos(\varphi_1-\varphi_2);\\
\tilde{\Gamma} &= \Gamma/\mu; \tau = t\mu/\pi R^2; \Omega_1 = \Omega\pi R^2/\mu
\end{aligned} \qquad \text{(B.7)}$$

**4. Case of boundary along the meridian.** When the solid boundary travels along a meridian with the coordinate $\varphi = \varphi_0$, we can obtain the following system for the sink of the vortex with coordinates $\theta_1 = \theta; \varphi_1 = \varphi$ (its mirror image coordinates are $\theta_2 = \theta_1 = \theta; \varphi_2 = 2\varphi_0 - \varphi_1 = 2\varphi_0 - \varphi$) from the system in Eq. (B.7):

$$\begin{aligned}
\frac{d\theta}{d\tau} &= -\frac{\tilde{\Gamma}\sin\theta\sin 2(\varphi_0-\varphi) + \sin\theta\cos\theta\left(1-\cos 2(\varphi_0-\varphi)\right)}{1-\sin^2\theta\left(1-\cos 2(\varphi_0-\varphi)\right)};\\
\frac{d\varphi}{d\tau} &= -\Omega_1 - \frac{\tilde{\Gamma}\cos\theta\left(1-\cos 2(\varphi-\varphi_0)\right) - \sin 2(\varphi_0-\varphi)}{1-\sin^2\theta\left(1-\cos 2(\varphi_0-\varphi)\right)};\\
\frac{d\varphi_2}{d\tau} &= -\Omega_1 + \frac{\tilde{\Gamma}\cos\theta\left(1-\cos 2(\varphi-\varphi_0)\right) - \sin 2(\varphi_0-\varphi)}{1-\sin^2\theta\left(1-\cos 2(\varphi_0-\varphi)\right)}
\end{aligned} \qquad \text{(B. 8)}$$

The system in Eq. (B.8) is self-consistent only for $\Omega_1 = 0$. In this case, the invariance of the sum of the longitudes of the helical vortices and its mirror hold for any instant: $\varphi + \varphi_2 = 2\varphi_0 = const$.

**5. Equations for the case of the same HVs.** From Eq. (B.4), the following system describing the interactions between the two identical anti-cyclonic point vortices combined with a point sink, where both are far from the solid boundary, can be obtained:

$$\begin{aligned}
\frac{d\theta_1}{d\tau} &= \frac{1}{A}\left(-\cos\theta_2\sin\theta_1 + \sin\theta_2\cos\theta_1\cos(\varphi_1-\varphi_2) + \tilde{\Gamma}\sin\theta_2\sin(\varphi_1-\varphi_2)\right);\\
\frac{d\theta_2}{d\tau} &= \frac{1}{A}\left(-\cos\theta_1\sin\theta_2 + \sin\theta_1\cos\theta_2\cos(\varphi_1-\varphi_2) - \tilde{\Gamma}\sin\theta_1\sin(\varphi_1-\varphi_2)\right);\\
\frac{d\varphi_1}{d\tau} &= -\Omega_1 + \frac{1}{A}\left(\tilde{\Gamma}\left(\cot\theta_1\sin\theta_2\cos(\varphi_1-\varphi_2) - \cos\theta_2\right) - \frac{\sin\theta_2\sin(\varphi_1-\varphi_2)}{\sin\theta_1}\right);\\
\frac{d\varphi_2}{d\tau} &= -\Omega_1 + \frac{1}{A}\left(\tilde{\Gamma}\left(\cot\theta_2\sin\theta_1\cos(\varphi_1-\varphi_2) - \cos\theta_1\right) + \frac{\sin\theta_1\sin(\varphi_1-\varphi_2)}{\sin\theta_2}\right)\\
A &= 1-\cos^2 u_{12}; \cos u_{12} = \cos\theta_1\cos\theta_2 + \sin\theta_1\sin\theta_2\cos(\varphi_1-\varphi_2);\\
\tilde{\Gamma} &= \Gamma/\mu; \tau = t\mu/\pi R^2; \Omega_1 = \Omega\pi R^2/\mu
\end{aligned} \tag{B.9}$$

The system obtained from the point helical cyclonic vortices is as follows:

$$\begin{aligned}
\frac{d\theta_1}{d\tau} &= \frac{1}{A}\left(-\cos\theta_2\sin\theta_1 + \sin\theta_2\cos\theta_1\cos(\varphi_1-\varphi_2) - \tilde{\Gamma}\sin\theta_2\sin(\varphi_1-\varphi_2)\right);\\
\frac{d\theta_2}{d\tau} &= \frac{1}{A}\left(-\cos\theta_1\sin\theta_2 + \sin\theta_1\cos\theta_2\cos(\varphi_1-\varphi_2) + \tilde{\Gamma}\sin\theta_1\sin(\varphi_1-\varphi_2)\right);\\
\frac{d\varphi_1}{d\tau} &= -\Omega_1 - \frac{1}{A}\left(\tilde{\Gamma}\left(\cot\theta_1\sin\theta_2\cos(\varphi_1-\varphi_2) - \cos\theta_2\right) + \frac{\sin\theta_2\sin(\varphi_1-\varphi_2)}{\sin\theta_1}\right);\\
\frac{d\varphi_2}{d\tau} &= -\Omega_1 - \frac{1}{A}\left(\tilde{\Gamma}\left(\cot\theta_2\sin\theta_1\cos(\varphi_1-\varphi_2) - \cos\theta_1\right) - \frac{\sin\theta_1\sin(\varphi_1-\varphi_2)}{\sin\theta_2}\right)\\
A &= 1-\cos^2 u_{12}; \cos u_{12} = \cos\theta_1\cos\theta_2 + \sin\theta_1\sin\theta_2\cos(\varphi_1-\varphi_2);\\
\tilde{\Gamma} &= \Gamma/\mu; \tau = t\mu/\pi R^2; \Omega_1 = \Omega\pi R^2/\mu
\end{aligned} \tag{B.10}$$

## Appendix C: Exact solution in the beta plane

To solve the AVCE for a rotating sphere in Eq. (1) according to an approximation in the beta plane, local Cartesian coordinates are typically introduced. They are expressed via spherical coordinates for a fixed radial coordinate that is equal to the radius of the sphere[5] (see (Sec. 7.7.9) in Ref. 5):

$$x = R\varphi \sin\theta_0 ; y = R(\theta_0 - \theta) \qquad \text{(C.1)}$$

To account for minor changes in the Coriolis parameter $f = 2\Omega\cos\theta$ with the latitude in Eq. (1), we use the so called beta-plane approximation in which the following representation of the Coriolis parameter is exploited (see (Sec. 7.7.11) in Ref. 5):

$$\begin{aligned} f &= f_0 + \beta y; \\ f_0 &= 2\Omega\cos\theta_0; \qquad \text{(C.2)} \\ \beta &= \frac{2\Omega}{R}\sin\theta_0 \end{aligned}$$

In Eq. (C.2), $\beta > 0$ in both hemispheres. For example, its estimated value for the middle latitude when the complement to the latitude is $\theta_0 = 45^0$ is $\beta = 1.62\times10^{-12} cm^{-1}\sec^{-1}$.[5]

Thus, the beta-plane approximation assumes that the effects of the spherical curvature of an extremely thin layer of fluid on the sphere, other than those in Eq. (C.2), can be neglected.

We assume that flow occurs in the layer rotating normal to the axis of the plane with an angular velocity that linearly changes in the north–south direction, defined by the positive direction of the *y*-axis. The positive $x$ -axis is directed from west to east.

Instead of the traditional Eq. (1), the following balance equation is used in beta-plane approximation:[75-79]

$$\begin{aligned} &\frac{\partial\omega}{\partial t} + u_x\frac{\partial\omega}{\partial x} + u_y\frac{\partial\omega}{\partial y} = 0; \\ &\omega = \omega_z + f_0 + \beta y; \\ &\omega_z \equiv \omega_r = -\left(\frac{\partial^2\psi}{\partial x^2} + \frac{\partial^2\psi}{\partial y^2}\right) \equiv -\Delta_2\psi; \qquad \text{(C.3)} \\ &u_x = \frac{dx}{dt} = \frac{\partial\psi}{\partial y}; u_y = \frac{dy}{dt} = -\frac{\partial\psi}{\partial x} \end{aligned}$$

The stream function follows directly from Eq. (C.3) as:

$$\frac{\partial \Delta_2 \psi}{\partial t} + \frac{\partial \psi}{\partial y}\frac{\partial \Delta_2 \psi}{\partial x} - \frac{\partial \psi}{\partial x}\frac{\partial \Delta_2 \psi}{\partial y} + \beta \frac{\partial \psi}{\partial x} = 0 \qquad \text{(C.4)}$$

Equation (C.4) corresponds to the usual form of the stream function in beta-plane approximation (see Refs. 75–79 and the references therein).

However, a special clarification is need for the applicability domain of this very form of the law of absolute vorticity conservation on a rotating sphere in the beta-plane approximation.

We can accurately derive in the limit $y / R << 1$ from the Eq. (1) and Eq. (C.1) the beta-plane approximation of the absolute vorticity conservation equation in the form:

$$\begin{aligned} &\frac{\partial \omega}{\partial t} + u_x \frac{\partial \omega}{\partial x} + u_y \frac{\partial \omega}{\partial y} = 0; \\ &\omega = \omega_z + f_0 + \beta y + O(y^2 / R^2); \\ &\omega_z \equiv \omega_r = -\left( \frac{\partial^2 \psi}{\partial x^2} + \frac{\partial^2 \psi}{\partial y^2} \right) + \frac{ctg\theta_0}{R}\frac{\partial \psi}{\partial y} \equiv -\Delta_2 \psi + \frac{ctg\theta_0}{R}\frac{\partial \psi}{\partial y}; \qquad \text{(C.5)} \\ &u_x = \frac{dx}{dt} = \frac{\partial \psi}{\partial y}; u_y = \frac{dy}{dt} = -\frac{\partial \psi}{\partial x} \end{aligned}$$

From Eq. (C.5), an equation for the stream function can be obtained that is different from Eq. (C.4):

$$\frac{\partial \Delta_2 \psi}{\partial t} + \frac{\partial \psi}{\partial y}\frac{\partial \Delta_2 \psi}{\partial x} - \frac{\partial \psi}{\partial x}\frac{\partial \Delta_2 \psi}{\partial y} + 2\beta \frac{\partial \psi}{\partial x} = \frac{ctg\theta_0}{R}\left( \frac{\partial^2 \psi}{\partial t \partial y} + \frac{\partial \psi}{\partial y}\frac{\partial^2 \psi}{\partial x \partial y} \right) \qquad \text{(C.6)}$$

Equation (C.6) coincides with Eq. (C.4)—for example, at the Equator, where the factor near the brace in the RHS of Eq. (C.6) becomes zero. Moreover, the domain of application of Eq. (C.4) can be obtained from the lower acceleration in the zonal direction compared with the zonal component of the Coriolis force, where this is acceptable when an exact balance between the Coriolis force and the force determined by the pressure gradient can be obtained, e.g., when the known condition of geostrophic balance holds. To verify this, it is necessary to compare the last term in the LHS of Eq. (C.6) with the sum of two terms in the parentheses on its RHS. The last condition always holds in the limit of a small Rossby parameter, which is the ratio of the local to the global vorticity due to the Earth's rotation. These assumptions are typically used to derive Eq. (C.3), but not by explicitly using Eq. (C.6).[5, 79] That is why in case of violations of the condition of geostrophic balance, it is necessary to either directly use Eq. (1) or its modified forms in beta-plane approximation—Eq. (C.5) or Eq. (C.6), but not Eq. (C.3) or Eq. (C.4)—for the intense vortices considered in this study as well as the dynamics of global vortices studied in Ref. 19.

An analog for the exact weak solution to Eq. (1) obtained in Ref. 19, when accurately accounting for the dependence of the Coriolis parameter on the latitude, is the solution to Eq. (C.5) (or Eq. (C.6)) as follows:

$$\psi = -\Omega R^2 \cos\theta_0 + A\frac{\Omega Rx}{\sin\theta_0} - \Omega Ry\sin\theta_0 + \Omega\frac{y^2}{2}\cos\theta_0 + \Omega\frac{y^3}{6R}\sin\theta_0 - \frac{1}{4\pi}\sum_{\alpha=1}^{N}\Gamma_\alpha \ln\left((x - x_\alpha(t))^2 + (y - y_\alpha(t))^2\right); \tag{C.7}$$

$$\omega = \sum_{\alpha=1}^{N}\Gamma_\alpha \delta(x - x_\alpha(t))\delta(y - y_\alpha(t)) \tag{C.8}$$

The value of the arbitrary constant of integration in Eq. (C.7) can be expressed via the intensity of the point source $\mu_0$ (or the intensity of the sink) as $A = \mu_0 / \pi R^2\Omega$ by considering the notation used in Eqs. (C.1) and (8).

Equation (C.7) of the stream function and Eq. (C.8) of the vortex field satisfy the weak exact solution to the AVCE in Eq. (C.5). To directly verify this, it is sufficient to substitute Eqs. (C.8) and (C.7) into Eq. (C.5), multiply the result by any finite function of coordinates, and integrate the product over all infinite planes of the variables (x, y) (see also Ref. 19 and the references therein).

When Eqs. (C.7) and (C.8) are considered in the context of the system of point HVs in beta-plane approximation, we can obtain a generalization to the case of the dynamics of point HVs, including the consideration of the radial component of velocity on the beta plane.

Then, by analogy with the field of velocity on a sphere potential in Eq. (10), we can obtain the following for the potential of the velocity field created by a system of N sources or sinks on the beta plane:[6, 7]

$$\Phi_2 = \frac{1}{4\pi}\sum_{\alpha=1}^{N}\mu_\alpha \ln\left((x - x_\alpha(t))^2 + (y - y_\alpha(t))^2\right) \tag{C.9}$$

$$V_x = \partial\Phi_2 / \partial x; V_y = \partial\Phi_2 / \partial y \qquad \text{(C.10)}$$

When Eqs. (C.9) and (C.10) are considered, it is necessary to substitute $u_x \to u_x + V_x; u_y \to u_y + V_y$ into Eq. (C.5), and to repeat the above procedure by substituting Eq. (C.8) into Eq. (C.5), as well as considering Eqs. (C.7), (C.9), and (C.10), to obtain the corresponding exact weak solution to it.

We can thus obtain an exact weak solution to Eq. (C.5) in the form of the following dynamic system for any number of strongly interacting point HVs on the beta plane:

$$U_\alpha \equiv dx_\alpha / dt = -\Omega R \sin\theta_0 \left(1 - \frac{y}{R} ctg\theta_0 - O\left(\left(\frac{y}{R}\right)^2\right)\right) + \sum_{\substack{\gamma=1 \\ \gamma\neq\alpha}}^{N} \frac{\mu_\gamma x_{\alpha\gamma} - \Gamma_\gamma y_{\alpha\gamma}}{2\pi b_{\alpha\gamma}}; \text{(C.11)}$$

$$V_\alpha \equiv dy_\alpha / dt = -\frac{\mu_0}{\pi R \sin\theta_0} + \sum_{\substack{\gamma=1 \\ \gamma\neq\alpha}}^{N} \frac{\mu_\gamma y_{\alpha\gamma} + \Gamma_\gamma x_{\alpha\gamma}}{2\pi b_{\alpha\gamma}}; \qquad \text{(C.12)}$$

$$b_{\alpha\gamma} = x_{\alpha\gamma}^2 + y_{\alpha\gamma}^2; x_{\alpha\gamma} = x_\alpha(t) - x_\gamma(t); y_{\alpha\gamma} = y_\alpha(t) - y_\gamma(t) \qquad \text{(C.13)}$$

For $\Omega = 0; \mu_0 = 0$, the system of Eqs. (C.11)–(C.13) exactly coincides with that given in Ref. 7 (see Sec. 3.1 in Ref. 7) for a dynamic system with an arbitrary number of the HVs on an infinite plane when it has invariants (Eq. (B.5)).

Note that Eq. (C.3) asymptotically coincides with the expressions considered in Refs. 75 and 76, relating to the potential vortices on the beta plane (see Eq. (1) in Ref. 75), and the balancing equation in the limit of an arbitrarily large deformation in the Rossby radius of $R_d \to \infty$. Equation (C.3) is identical to the expressions in Refs. 75 and 76 when $\Delta_2\psi \to \Delta_2\psi - \frac{1}{R_d^2}\psi$ is substituted into it.

We now consider the exact solutions of Eqs. (C.11) and (C.12) for the case of a single vortex on the beta plane:

$$U = U_1 = -\Omega R \sin\theta_0 \left(1 - \frac{y}{R} ctg\theta_0\right);$$
$$V = V_1 = -\frac{\mu_0}{\pi R \sin\theta_0}; y = V_1 t + y(0) \quad \text{(C.14)}$$

To obtain the individual velocities of transfer of finite vortices along a latitude to the west as well as the velocity of the vortex system as a whole, it is sufficient to have the finite Coriolis parameter $f_0 \neq 0$ even if the beta effect is not considered when $\beta = 0$.

Note that in the limit of $A_m / R_d \to 0$, where $A_m$ is the effective monopole radius of the vortex, in Ref. 76, the following component of velocity of the monopole is estimated on the beta plane along the $x$-axis (see Eqs. (64) and (65) in Ref. 76):

$$U = -\beta \frac{A_m^2}{4} = -\frac{\Omega A_m^2 \sin\theta_0}{2R} \quad \text{(C.16)}$$

We obtained an exact solution to the AVCE in Eq. (1) in this study by considering a general form of the rotation of a sphere by introducing the stream function $\psi_0$ of regular flow in a superposition with the stream function of a system of point vortices. We were thus able to precisely compensate for the global vortical field. As a result, the field of absolute vorticity in Eq. (1) is described by only a singular component. This helps avoid having to solve a complicated system of integro-differential equations, as is the norm in the traditional treatment of the interactions of a field of singular vortices with the field of a regular vortex that has emerged due to the rotation of the sphere.[75, 77]

The work presented here advances approaches for treating HVs on a rotating sphere as well as on the beta plane. The exact solutions provided here were considered in the relation to the modeling of the dynamics of a TC.

## Data Availability

The data that support the findings of this study are available within the article.